\documentclass[twocolumn]{aastex631}

\usepackage{lineno}
\usepackage{graphicx}
\usepackage{longtable}
\usepackage{bm}
\usepackage{booktabs}
\usepackage{multirow}
\usepackage{float}
\usepackage{hyperref}
\usepackage{url}
\usepackage{wrapfig}
\definecolor{crimson}{RGB}{230, 20, 40}

\accepted{October 14, 2023}

\submitjournal{ApJ}

\shorttitle{Outflows in NGC 3256}
\shortauthors{Bohn et al.}

\graphicspath{{./}{figures/}}

\begin{document}

\title{GOALS-JWST: The Warm Molecular Outflows of the Merging Starburst Galaxy NGC 3256}

\email{tbohn002@ucr.edu}

\author[0000-0002-4375-254X]{Thomas Bohn}
\affil{Hiroshima Astrophysical Science Center, Hiroshima University, 1-3-1 Kagamiyama, Higashi-Hiroshima, Hiroshima 739-8526, Japan}

\author[0000-0003-4268-0393]{Hanae Inami}
\affiliation{Hiroshima Astrophysical Science Center, Hiroshima University, 1-3-1 Kagamiyama, Higashi-Hiroshima, Hiroshima 739-8526, Japan}

\author[0000-0001-5042-3421]{Aditya Togi}
\affiliation{Department of Physics, Texas State University, 601 University Drive, San Marcos, TX 78666, USA}

\author[0000-0003-3498-2973]{Lee Armus}
\affiliation{IPAC, California Institute of Technology, 1200 E. California Blvd., Pasadena, CA 91125}

\author[0000-0001-8490-6632]{Thomas S.-Y. Lai}
\affil{IPAC, California Institute of Technology, 1200 E. California Blvd., Pasadena, CA 91125}

\author[0000-0003-0057-8892]{Loreto Barcos-Munoz}
\affiliation{National Radio Astronomy Observatory, 520 Edgemont Road, Charlottesville, VA 22903, USA}
\affiliation{Department of Astronomy, University of Virginia, 530 McCormick Road, Charlottesville, VA 22903, USA}

\author[0000-0002-3139-3041]{Yiqing Song}
\affiliation{European Southern Observatory, Alonso de Córdova, 3107, Vitacura, Santiago, 763-0355, Chile}
\affiliation{Joint ALMA Observatory, Alonso de Córdova, 3107, Vitacura, Santiago, 763-0355, Chile}

\author[0000-0002-1000-6081]{S. T. Linden}
\affiliation{Department of Astronomy, University of Massachusetts at Amherst, Amherst, MA 01003, USA}

\author[0000-0001-7291-0087]{Jason Surace}
\affiliation{IPAC, California Institute of Technology, 1200 E. California Blvd., Pasadena, CA 91125}

\author[0000-0002-6570-9446]{Marina Bianchin}
\affiliation{Department of Physics and Astronomy, 4129 Frederick Reines Hall, University of California, Irvine, CA 92697, USA}

\author[0000-0002-1912-0024]{Vivian U}
\affiliation{Department of Physics and Astronomy, 4129 Frederick Reines Hall, University of California, Irvine, CA 92697, USA}

\author[0000-0003-2638-1334]{Aaron S. Evans}
\affiliation{National Radio Astronomy Observatory, 520 Edgemont Road, Charlottesville, VA 22903, USA}
\affiliation{Department of Astronomy, University of Virginia, 530 McCormick Road, Charlottesville, VA 22903, USA}

\author[0000-0002-5666-7782]{Torsten B\"oker}
\affiliation{European Space Agency, Space Telescope Science Institute, Baltimore, MD 21218, USA}

\author[0000-0001-6919-1237]{Matthew A. Malkan}
\affiliation{Department of Physics \& Astronomy, UCLA, Los Angeles, CA 90095-1547}

\author[0000-0003-3917-6460]{Kirsten L. Larson}
\affiliation{AURA for the European Space Agency (ESA), Space Telescope Science Institute, 3700 San Martin Drive, Baltimore, MD 21218, USA}

\author[0000-0002-2596-8531]{Sabrina Stierwalt}
\affiliation{Occidental College, Physics Department, 1600 Campus Road, Los Angeles, CA 90042}

\author[0009-0003-4835-2435]{Victorine A. Buiten}
\affiliation{Leiden Observatory, Leiden University, PO Box 9513, 2300 RA Leiden, The Netherlands}

\author[0000-0002-2688-1956]{Vassilis Charmandaris}
\affiliation{Department of Physics, University of Crete, Heraklion, 71003, Greece}
\affiliation{Institute of Astrophysics, Foundation for Research and Technology-Hellas (FORTH), Heraklion, 70013, Greece}
\affiliation{School of Sciences, European University Cyprus, Diogenes street, Engomi, 1516 Nicosia, Cyprus}

\author[0000-0003-0699-6083]{Tanio Diaz-Santos}
\affiliation{Institute of Astrophysics, Foundation for Research and Technology-Hellas (FORTH), Heraklion, 70013, Greece}
\affiliation{School of Sciences, European University Cyprus, Diogenes street, Engomi, 1516 Nicosia, Cyprus}

\author[0000-0001-6028-8059]{Justin H. Howell}
\affiliation{IPAC, California Institute of Technology, 1200 E. California Blvd., Pasadena, CA 91125}

\author[0000-0003-3474-1125]{George C. Privon}
\affiliation{National Radio Astronomy Observatory, 520 Edgemont Road, Charlottesville, VA 22903, USA}
\affiliation{Department of Astronomy, University of Virginia, 530 McCormick Road, Charlottesville, VA 22903, USA}
\affiliation{Department of Astronomy, University of Florida, P.O. Box 112055, Gainesville, FL 32611, USA}

\author[0000-0001-5231-2645]{Claudio Ricci}
\affiliation{Instituto de Estudios Astrof\'isicos, Facultad de Ingenier\'ia y Ciencias, Universidad Diego Portales, Av. Ej\'ercito Libertador 441, Santiago, Chile}
\affiliation{Kavli Institute for Astronomy and Astrophysics, Peking University, Beijing 100871, China}

\author[0000-0001-5434-5942]{Paul P. van der Werf}
\affiliation{Leiden Observatory, Leiden University, PO Box 9513, 2300 RA Leiden, The Netherlands}

\author[0000-0002-5828-7660]{Susanne Aalto}
\affiliation{Department of Space, Earth and Environment, Chalmers University of Technology, 412 96 Gothenburg, Sweden}

\author[0000-0003-4073-3236]{Christopher C. Hayward}
\affiliation{Center for Computational Astrophysics, Flatiron Institute,
162 Fifth Avenue, New York, NY 10010, USA}

\author[0000-0002-6650-3757]{Justin A. Kader}
\affiliation{Department of Physics and Astronomy, 4129 Frederick Reines Hall, University of California, Irvine, CA 92697, USA}

\author[0000-0002-8204-8619]{Joseph M. Mazzarella}
\affiliation{IPAC, California Institute of Technology, 1200 E. California Blvd., Pasadena, CA 91125}

\author[0000-0002-2713-0628]{Francisco Muller-Sanchez}
\affiliation{Department of Physics and Materials Science, The University of Memphis, 3720 Alumni Avenue, Memphis, TN 38152, USA}

\author[0000-0002-1233-9998]{David B. Sanders}
\affiliation{Institute for Astronomy, University of Hawaii, 2680 Woodlawn Drive, Honolulu, HI 96822, USA}

\begin{abstract}

We present \textit{James Webb Space Telescope (JWST)} Integral Field Spectrograph observations of NGC 3256, a local infrared-luminous late-stage merging system with two nuclei roughly 1$\;\rm{kpc}$ apart, both of which have evidence of cold molecular outflows. Using \textit{JWST}/NIRSpec and MIRI datasets, we investigate this morphologically complex system on spatial scales of $<$100$\;\rm{pc}$, where we focus on the warm molecular H$_2$ gas surrounding the nuclei. We detect collimated outflowing warm H$_2$ gas originating from the southern nucleus, though we do not find significant outflowing H$_2$ gas surrounding the northern nucleus. We measure maximum intrinsic outflow velocities of $\sim$1,000$\;\rm{km}\;\rm{s}^{-1}$, which extend out to a distance of 0.7$\;\rm{kpc}$. Based on H$_2$ S(7)/S(1) ratios, we find a larger fraction of warmer gas near the S nucleus, which decreases with increasing distance from the nucleus, signifying the southern nucleus as a primary source of H$_2$ heating. The gas mass of the warm H$_2$ outflow component is estimated to be $M\rm{_{warm,out}}=(1.4\pm0.2)\times10^6\;\rm{M}_{\odot}$, as much as 6$\%$ of the cold H$_2$ mass estimated using ALMA CO data. The outflow time scale is about $7\times10^5\;\rm{yr}$, resulting in a mass outflow rate $\dot{M}\rm{_{warm,out}}=2.0\pm0.8\;\rm{M}_{\odot}\;\rm{yr}^{-1}$ and kinetic power $P\rm{_{warm,out}}\;\sim\;4\times10^{41}\;\rm{erg}\;\rm{s}^{-1}$. Lastly, regions within our $3\farcs0\times3\farcs0$ NIRSpec data where the outflowing gas reside show high [\ion{Fe}{2}]/Pa$\beta$ and H$_2$/Br$\gamma$ line ratios, indicate enhanced mechanical heating caused by the outflows. The fluxes and ratios of Polycyclic Aromatic Hydrocarbons (PAH) in these regions are not significantly different compared to those elsewhere in the disk, suggesting the outflows may not significantly alter the PAH ionization state or grain size.

\end{abstract}


\section{Introduction} \label{sec:intro}

The physical processes that govern the evolution of galaxies are of fundamental importance to our understanding of how galaxies in the Universe formed. These processes manifest in many forms, including internal mechanisms, such as stellar and active galactic nucleus (AGN) outflows \citep[e.g.,][]{Muller2011,U2019,Aravindan2023}, and external processes such as galaxy mergers \citep[e.g.,][]{Lin2008,Privon2013,Conselice2014}. Outflows are believed to regulate and suppress star formation and can, in many instances, lead their hosts to the well-defined red sequence \citep[e.g.,][]{Croton2006,King2015,Veilleux2020}. Major mergers also play a critical role in the formation of massive galaxies, and are attributed to the coevolution of the host galaxy and supermassive black holes \citep[e.g.,][]{Medling2015,Ricci2017b,Ramos2019} and the formation of compact star clusters \citep[e.g.,][]{Mulia2015,Adamo2020,Linden2021}. Local luminous infrared galaxies (LIRGs; $L_{\rm IR}>10^{11}L_{\odot}$) are ideal systems for studying these processes because they allow us to analyze evolutionary processes across the full array of merger states, and evaluate the elevated star formation rates triggered by such mergers. Their diversity in starburst and AGN power, merger stages, and outflow activity makes LIRGs compelling ecosystems for tying these properties to the stages of galaxy evolution.

With an infrared luminosity of $L_{\rm IR,\; 8-1000\mu m}=3.6\times10^{11}\;L_{\odot}$ and redshift of $z=0.0094$, NGC 3256 is the most luminous system within $z=0.01$ \citep{Sanders2003}. This system, composed of two gas-rich disk galaxies in a late-stage major merger, is part of the Great Observatories All-sky LIRGs Survey \citep[GOALS;][]{Armus2009}, a sample of bright ($S_{60}>$ 5.24 Jy) LIRGs selected from the \textit{IRAS} Revised Bright Galaxy Sample \citep{Sanders2003}. The system shows a complex and tidally disturbed morphology, with significantly distorted spiral arms and prominent dust lanes. The two nuclei are in close proximity to each other, $5\farcs0$ ($\sim$1 kpc), and this system has been extensively observed from X-ray to radio wavelengths \citep{Rich2011,Piqueras2012,Stierwalt2013,Sakamoto2014,Ohyama2015,Chisholm2016,Harada2018}. These observations have revealed an optically unobscured northern (N) nucleus with signs of starburst activity \citep{Doyon1994,Boeker1997,Lira2002,Lipari2004,Pereira2011,Laha2018}, while the southern (S) nucleus has been described as a heavily obscured, low-luminosity or nascent AGN \citep{Neff2003,Sakamoto2014,Emonts2014,Ohyama2015,Michiyama2018,Yamada2021}.

One of the most compelling aspects of NGC 3256 is the presence of outflows in both nuclei. Submillimeter ALMA observations indicate two different biconical outflows of molecular CO gas originating from the nuclei, with deprojected maximum velocities (and inclination angles) of $\sim$750 (60$^\circ$) and 2600 km s$^{-1}$ (20$^\circ$) for the outflows in the northern and southern galaxies, respectively \citep{Sakamoto2014,Michiyama2018}. Near-infrared (NIR) VLT/SINFONI observations of the S nucleus also reveal biconical outflows in the warm, $>$2000 K H$_2$ gas, where the gas mass of these outflowing components is estimated to be $\sim$1,200 $M_{\odot}$ \citep{Emonts2014}. 



While previous observations have been able to analyze the warm and cold molecular gas phases in NIR and submillimeter wavelengths with great detail, analysis of the warm H$_2$ component in the mid-infrared (MIR) has been mainly restricted to using the S(1)--S(3) transitions due to the resolution and sensitivity of the \textit{Spitzer Space Telescope} \citep[e.g.][]{Stierwalt2013,Inami2013}. \cite{Petric2018} calculated the mass of the warm molecular H$_2$ component of the northern galaxy to be log($M\rm{_{warm}}/\rm{M_{\odot}}$) = 7.38 and a temperature of 322 K within the IRS slits that were centered on the N nucleus. While \textit{Spitzer} gave a great first look at the warm H$_2$ gas, the H$_2$ emission was not spatially resolved and the spectral resolution of \textit{Spitzer}, R$\sim$60--600, made it difficult to spectrally resolve the outflowing gas in NGC 3256.

The various H$_2$ transitions and their line ratios with other IR lines are useful for identifying the sources heating the molecular gas and energizing the interstellar medium \citep[ISM;][]{Larkin1998,Riffel2013,Colina2015}. They can be used to identify the presence of shocks induced by the outflows, and help evaluate the role outflows have in regulating star formation. With \textit{JWST}, we can now perform the most detailed, spatially resolved analysis of outflowing warm H$_2$ gas to date in NGC 3256. The unprecedented spatial resolution of \textit{JWST} enables us to examine the warm H$_2$ component on scales of $\sim$40--100 pc. We also take advantage of the superb sensitivity and wavelength coverage of \textit{JWST} to analyze the H$_2$ 0--0 S(1)--S(8) rotational lines between 5--17 $\mu$m. Specifically, we utilize integral-field spectroscopy (IFS) of the medium-resolution spectrometer (MRS) of the Mid-Infrared Instrument \citep[MIRI,][]{Rieke2015,Labiano2021,Wright2023} and the Near-InfraRed Spectrograph \citep[NIRSpec,][]{Boker2022,Jakobsen2022} to assess the kinematics and energetics of the outflows, and evaluate their impact on the local ISM. 

In our analysis, we adopt a cosmology of $H_0=70$ km s$^{-1}$ Mpc$^{-1}$, $\Omega_\Lambda=0.72$, and $\Omega_{\rm matter}=0.28$. At the redshift of NGC 3256, $z=0.0094$ (40.4 Mpc), $1\farcs0$ subtends $\sim$190 pc.


\begin{figure*}
\epsscale{1.2}
\plotone{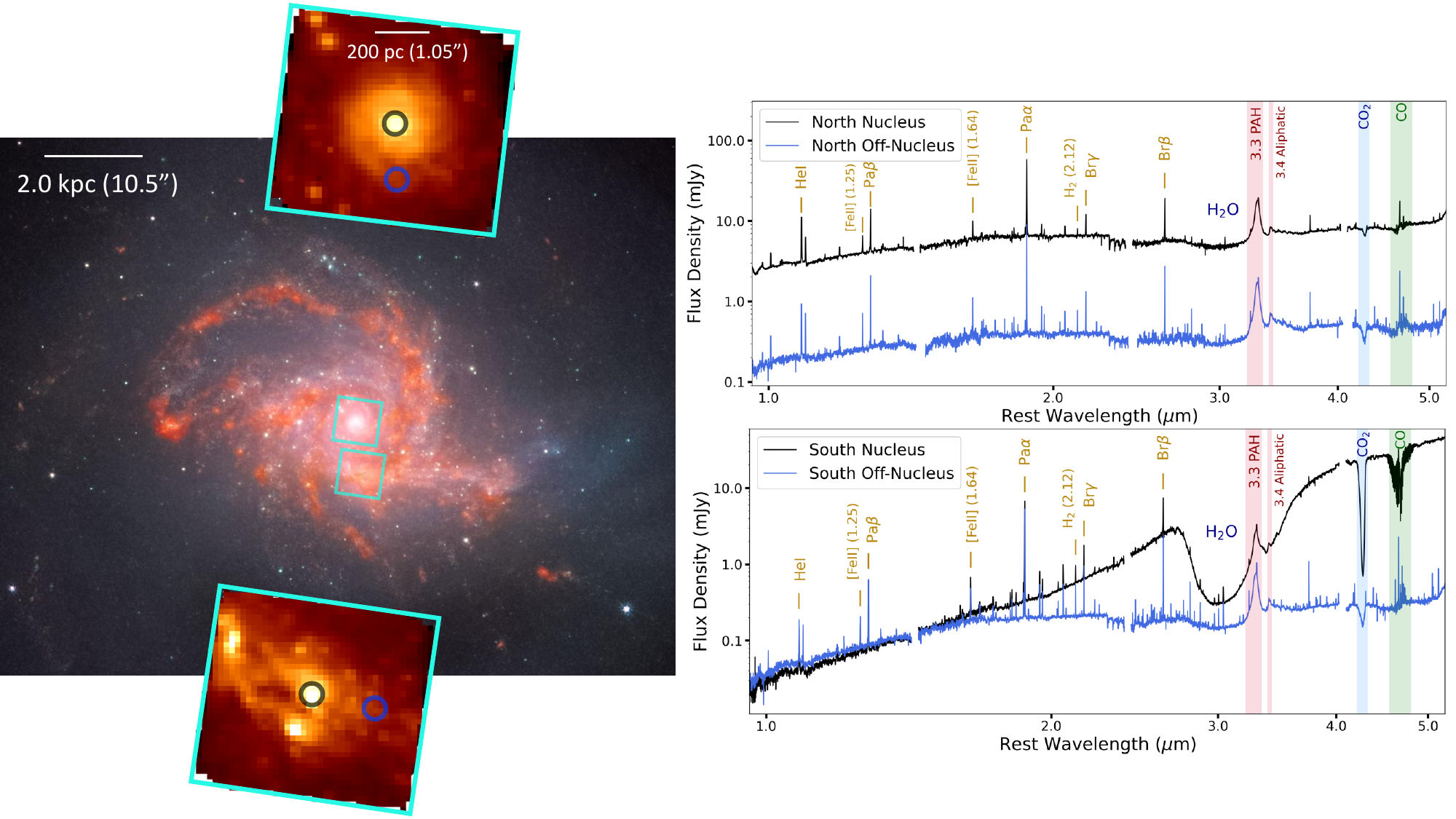}
\caption{(\textit{middle-left}) \textit{JWST}/NIRCam false-color image of NGC 3256 using F150W (1.5 $\mu$m), F200W (2.0 $\mu$m), and F444W (4.4 $\mu$m) filters. (\textit{top-left, bottom-left}) NIRSpec G235H/F170LP 2 $\mu$m continuum images of each nucleus are also shown (field of view size). These images correspond to the locations indicated by the cyan boxes in the NIRCam image. (\textit{right}) Full NIRSpec spectra (0.97--5.27 $\mu$m) which were extracted from the nuclear (black) and off-nuclear (blue) regions using a $0\farcs2$ radius aperture (top for the northern nucleus and bottom for the southern nucleus). Their extraction locations and sizes are indicated with the same color in the top-left and bottom-left images, and the spectra shown here have been corrected for aperture losses. The southern nucleus is heavily obscured, significantly redder in the MIR, and has strong 3.05 $\mu$m H$_2$O and 4.27 $\mu$m CO$_2$ ice absorption. \label{fig:RGB_plot}}
\end{figure*}

\section{Observations and Data Reduction} \label{sec:observations}

Observations of NGC 3256 were taken as part of the Early Release Science (ERS) Program 1328 (co-PIs: L. Armus and A. Evans), for which the \textit{JWST}/NIRSpec IFS data were taken on 2022 December 23-24. Two sets of observations, each centered on a nucleus, were taken using three high-resolution gratings: G140H/F100LP, G235H/F170LP, and G395H/F290LP. This resulted in a wavelength coverage of 0.97--5.27 $\mu$m, with a nominal resolving power of R$\sim$2700. A four-point dither pattern was utilized, and the total exposure times for all gratings were 934 s and 2218 s for the northern and southern pointings, respectively. Observations using \textit{JWST}/MIRI in the MRS mode were taken on 2022 December 27. Like for NIRSpec, a separate pointing for each nucleus was taken. These observations include all three subbands (short [A], medium [B], and long [C]) in all four channels, which cover the full wavelength range from 4.9--28.8 $\mu$m. The total exposure time for each subband/channel was 444 s, and was the same for both pointings. In this article, we predominately use channels 1--3, which have a nominal resolving power of R$\sim$2000--3700.


The data were processed through the standard \textit{JWST} Calibration Pipeline \citep[ver.1.12.5 for NIRSpec; ver.1.11.3 for MIRI,][]{Bushouse2023}. Stage 1 of the reduction pipeline, \textsc{Detector1}, implements detector-level corrections and outputs count-rate data products. Stage 2 (\textsc{spec2}) applies various calibrations steps, including distortion, wavelength, and flux calibrations. Stage 3 (\textsc{spec3}) performs background subtraction for MIRI data and builds the final data cubes from the individual exposures. We also incorporated leakcal calibration in our NIRSpec data. Before stage 3 was implemented, we only included spaxels with data quality (DQ) equal to 0 in order to remove bad spaxels contaminating the NIRSpec data products. We additionally flagged any spaxels in the postpipeline products with values significantly above emission lines or that were negative.

To properly align both the NIRSpec and MIRI data, Gaia DR3 \citep{Gaia2016,Gaia2023} reference frames of nearby field stars were first used to derive WCS corrections for the \textit{JWST}/NIRCam and MIRI images \citep[also obtained by ERS 1328, see][]{Linden2024}. The IFS cubes were subsequently collapsed into a single frame and aligned to the WCS corrected NIRCam/MIRI images, where corrections were R.A.: +$0\farcs02$ and decl.: +$0\farcs32$ for NIRSpec, and R.A.: $-0\farcs01$ and decl.: +$0\farcs12$ for MIRI. 

Lastly, fringing corrections were applied to all extracted 1D MIRI spectra using the calibration pipeline. We also stitched the spectra by scaling all the continua to that of MIRI channel 3 long, the longest wavelength channel used in our analysis, where scaling factors were typically around 5$\%$. 


\section{Methods of Analysis} \label{sec:Analysis}

The primary focus of this article is to obtain robust mass estimates and temperatures of the warm outflowing H$_2$ gas, as well as analyze its kinematics and energetics. Due to the complex kinematic nature of this merging system, robust emission-line decompositions and detailed mapping of the spatial structures are needed to accurately identify and study the H$_2$ gas. To accomplish this, we used the Bayesian AGN Decomposition Analysis for SDSS Spectra \citep[\textsc{BADASS}\footnote{\url{https://github.com/remingtonsexton/BADASS3}},][]{Sexton2021} code. Through Markov-Chain Monte Carlo (MCMC) routines, \textsc{BADASS} performs simultaneous multicomponent fits to emission-line spectra. Individual spaxels in both the NIRSpec and MIRI datasets were fit iteratively across the full spatial extent of the data cubes, where a third-order Legendre polynomial was used for the continuum. Emission lines were fit with either Gaussian (recombination, molecular, and fine-structure lines) or Lorentzian functions (polycyclic aromatic hydrocarbon (PAH) features). A secondary Gaussian component to the fit was included based on the \textit{F}-test of variances: \textit{F} = $(\sigma\rm{_{single}})^{2}/(\sigma\rm{_{double}})^{2}$, where $\sigma$ is the standard deviation of the residuals using either a single or double Gaussian fit. If \textit{F} $>$ 3.0, then adding a second component provided a significant improvement to the fit and was thus justifiable. This secondary component was restricted to have a larger velocity offset from the rest-frame wavelength than the main component. Due to the diverse emission profiles in our data, we left the amplitude and width as unrestricted free parameters. We also set a signal-to-noise ratio (S/N) threshold of five for all lines and components to determine a detection. Uncertainties were derived from the error spectrum and the random errors associated with the Bayesian fitting process. 



\begin{figure*}
\epsscale{1.15}
\centering
\plotone{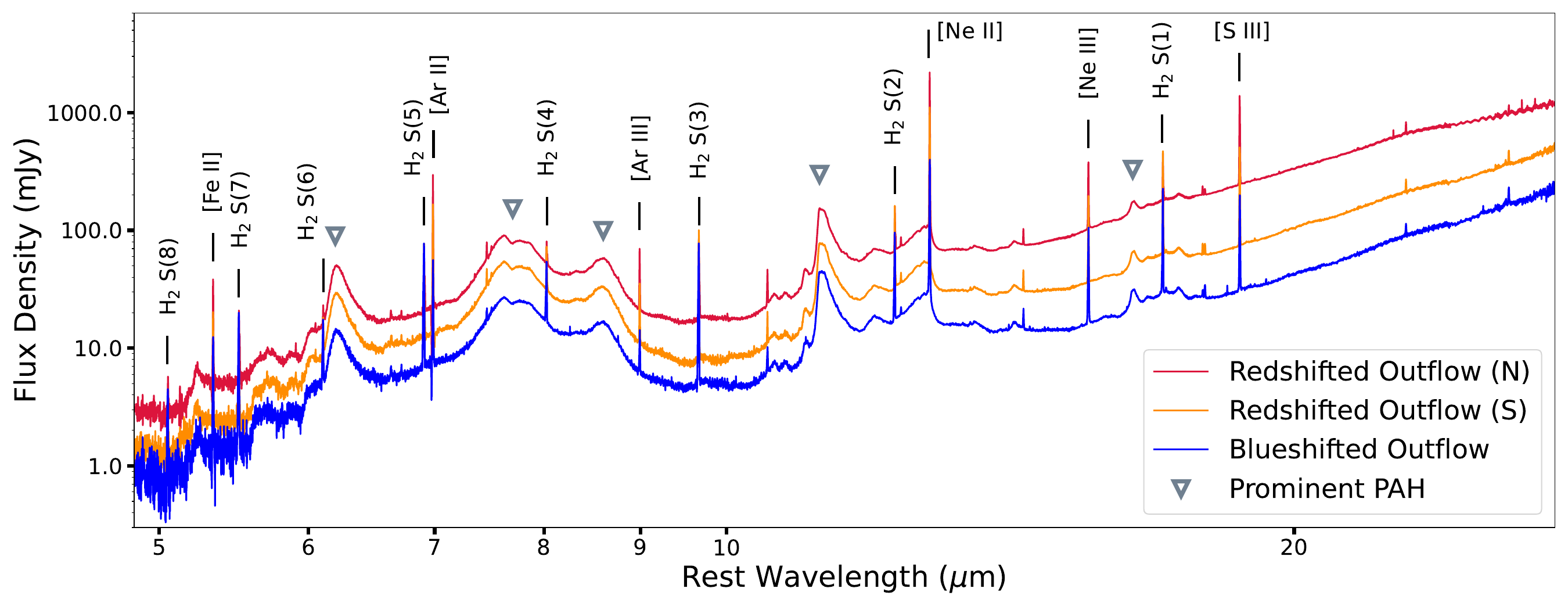}
\caption{Full MIRI spectra (4.9--27.9 $\mu$m), extracted from the regions where we detect emission from the secondary (outflow) component seen in all MIRI channels, about 350 pc to the south (blue), and 350 pc (orange) and 650 pc (red) to the north of the S nucleus (see dotted circles in Figure \ref{fig:Flux_maps}). An aperture radius of r=$0\farcs8$ was used, and the spectra here have been aperture corrected but are not corrected for extinction. The rotational H$_2$ lines used in this study are labeled, along with other MIR emission lines. Prominent PAH features are also marked with upside-down triangles. While the emission features of both spectra are similar, the continuum of the blueshifted outflow is as much as 3 times lower than that of the redshifted outflow for regions equidistant from the nucleus. \label{fig:Full_MIRI}}
\end{figure*}

\section{Results} \label{sec:Results}

\subsection{Near-infrared Spectra} \label{sec:NIRSpec_spectra}

In Figure \ref{fig:RGB_plot}, we show NIRCam images and NIRSpec spectra of the two nuclei and two nearby regions in the disks of NGC 3256. The spectra shown were extracted using an aperture of radius $0\farcs2$. In the nuclear and off-nuclear regions of both galaxies, we detect a wealth of hydrogen recombination, H$_2$, \ion{He}{1}, and [\ion{Fe}{2}] emission lines. 
 
The spectrum of the N nucleus is similar to that of the off-nuclear regions. They both show a flat continuum and weak H$_2$O (3.05 $\mu$m) and CO$_2$ (4.27 $\mu$m) ice absorption. The PAH emission feature at 3.3 $\mu$m is also clearly seen. In contrast, the S nucleus shows a rising MIR continuum and deep H$_2$O and CO$_2$ ice absorption. The ice absorption in the S nucleus is significant, with the continuum level dropping by over an order of magnitude. There is also pronounced absorption in the CO rovibrational band at 4.67 $\mu$m \citep{Pereira2024}. 

Although the NIRSpec observations do not cover the full extent of the CO outflow regions as detected by ALMA \citep{Sakamoto2014,Michiyama2018}, the regions where there is overlap show blue and red asymmetric profiles in many molecular and ionized emission lines, including [\ion{Fe}{2}] 1.25 $\mu$m, H$_2$ 1--0 S(3) 1.96 $\mu$m, and H$_2$ 1--0 S(1) 2.12 $\mu$m, that are not seen elsewhere. These blue and red emission wings are detected to within 100 pc of the S nucleus and extend beyond the NIRSpec FOV. In order to study the full extent of the H$_2$ emission, much of our analysis will focus on the MIRI data, which cover a larger area than NIRSpec.

\begin{figure*}
\epsscale{1.15}
\centering
\plotone{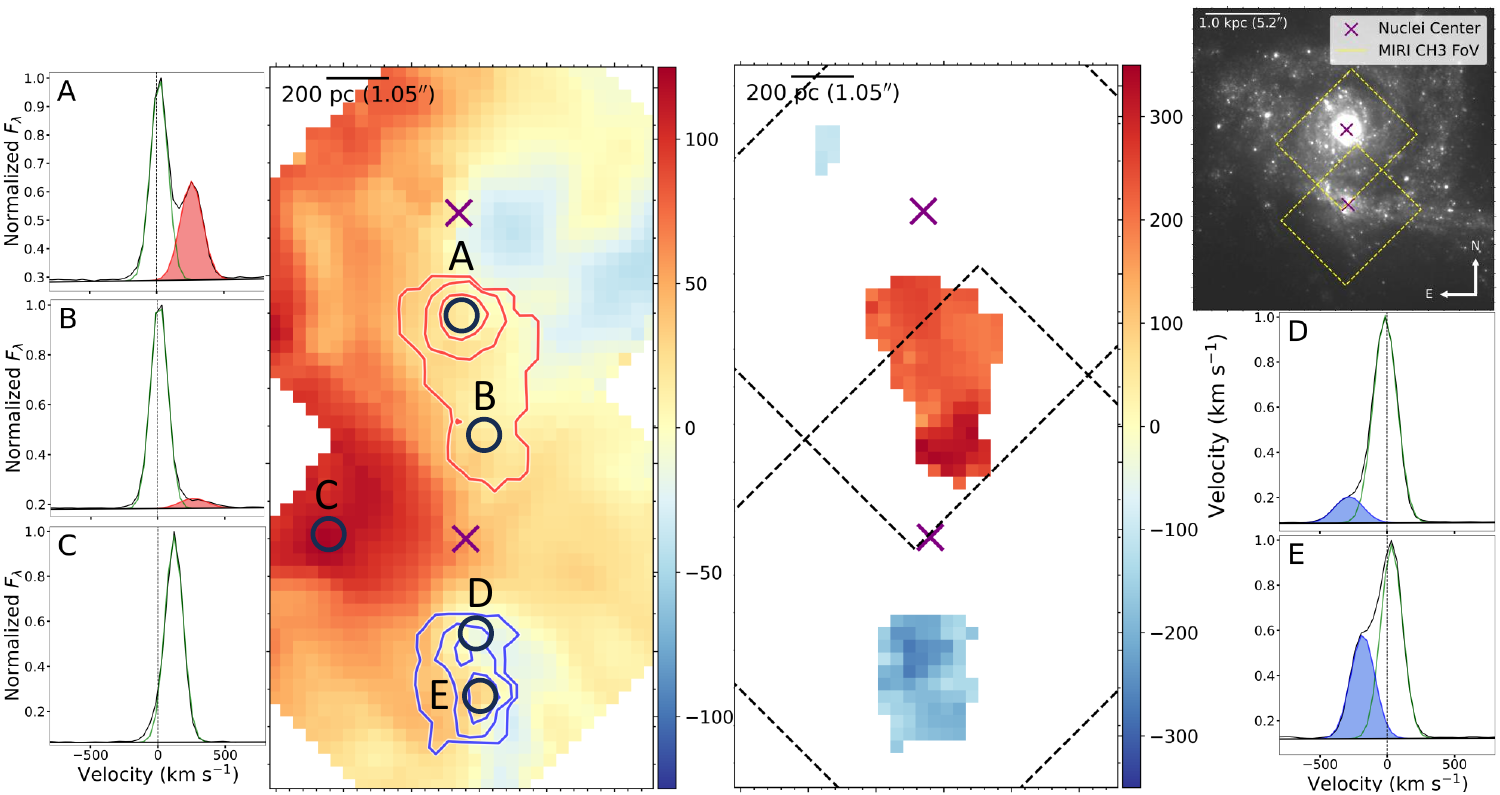}
\caption{(\textit{center}) H$_2$ S(1) 17.0 $\mu$m velocity maps (in units of km s$^{-1}$) of the main and secondary (outflow) gas components, where the main component is defined as the emission profile closest to the systemic velocity. Flux contours (1.0, 8.0, and 10.0$\times$10$^{-20}$ W m$^{-2}$) of the H$_2$ S(1) outflow component are overplotted in red and blue, and the purple crosses are the locations of the nuclei. The full spatial extent of the outflows based on our detection thresholds are plotted in the central-right velocity map (with an adjusted color scale). (\textit{upper-right}) NIRCam F200W image with the spatial coverage of the velocity map (MIRI channel 3 FoV) is overplotted. (\textit{left/right}) Aperture extractions (r=$0\farcs4$) of the normalized S(1) emission at various locations (A - E) are overplotted in the velocity map of the main component and show the diverse emission profiles. The vertical dotted line represents the rest-frame wavelength of S(1). The best-fit model is plotted for each extraction, where the green line represents the main component fit, and the blue and red filled Gaussians represent the blueshifted and redshifted outflow components, respectively. The majority of the outflowing components are spectrally resolved from the main component and show the structure extending to the north and south from the southern nucleus. \label{fig:Vel_maps}}
\end{figure*}

\begin{figure*}
\epsscale{1.15}
\centering
\plotone{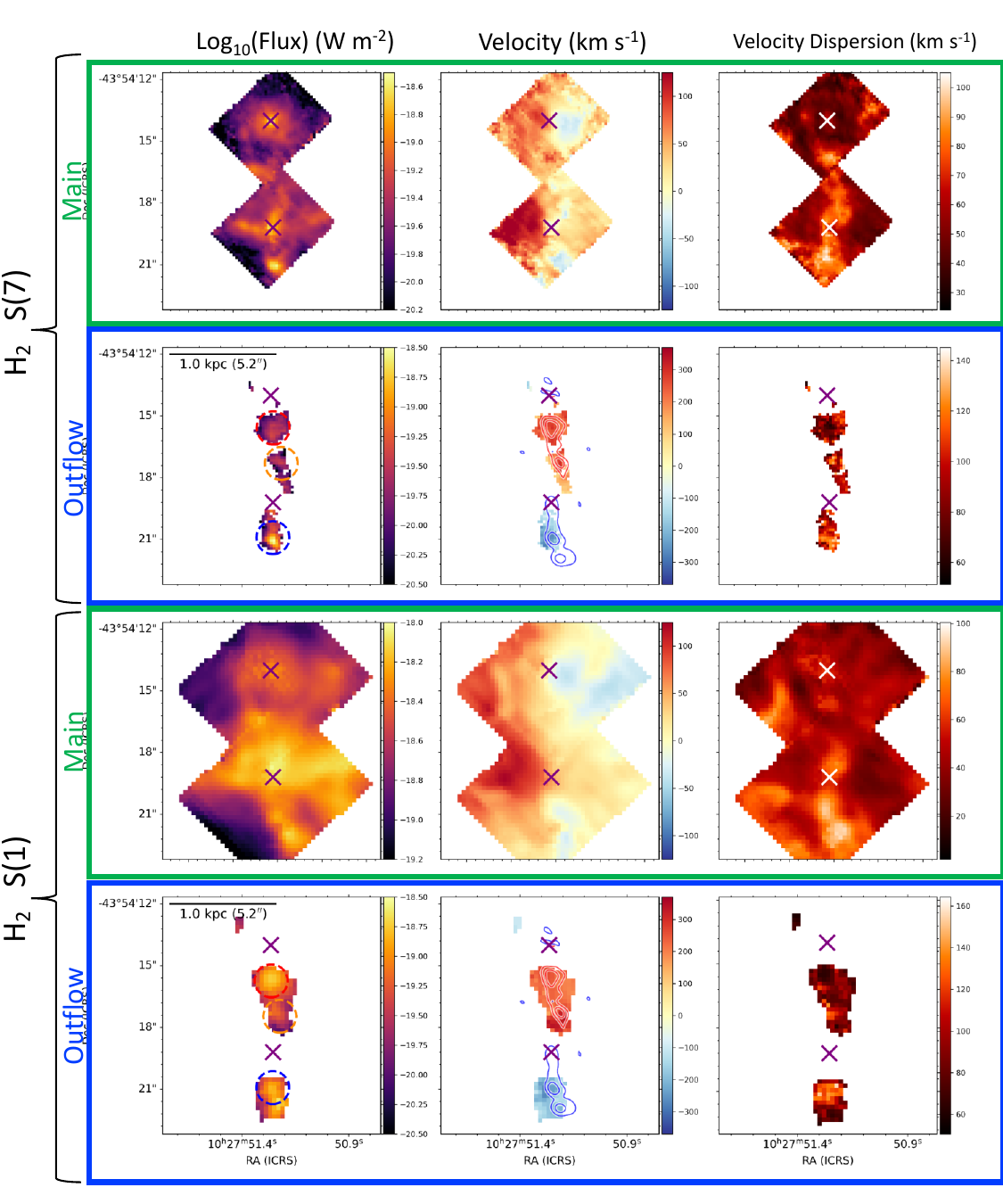}
\caption{(\textit{left to right}) Flux, velocity, and dispersion maps of H$_2$ S(7) (top two rows) and S(1) (bottom two rows) emission lines. Both the main and secondary (outflow) components (see Section \ref{sec:Analysis}) are plotted separately. The locations of the nuclei are shown as purple (or white) crosses. The r=$0\farcs8$ apertures used to calculate the H$_2$ gas masses are shown as dotted red, orange, and blue circles in the outflow flux panels. ALMA CO flux contours, with velocities set at $\pm$300 km s$^{-1}$ \citep{Sakamoto2014}, are overplotted as blue and red contours in the outflow velocity (central) panels. The outflowing warm molecular gas is collimated, extends out to a projected distance of $\sim$600 pc from the S nucleus, and is cospatial with the cold molecular outflow. \label{fig:Flux_maps}}
\end{figure*}

\subsection{Warm H$_2$ Gas} \label{sec:Molecular_Gas}

The wavelength coverage of \textit{JWST}/MIRI allows us to examine the warm H$_2$ gas using the H$_2$ 0--0 S(8)--S(1) emission lines from 5.05 to 17.0 $\mu$m. The larger spatial coverage of MIRI (compared to NIRSpec) allows us to better trace the extent of the H$_2$ emission. In Figure \ref{fig:Full_MIRI}, we plot the full MIRI spectra extracted from three regions where we detect emission of the secondary, asymmetric H$_2$ component observed in all MIRI channels. One aperture was placed 350 pc south of the S nucleus where we measure strong blueshifted gas emission, while two other apertures were placed 350 and 650 pc north of the S nucleus where we detect redshifted gas emission. The continuum and emission features of all spectra are similar, albeit the continuum level of the redshifted gas is as much as 3 times higher than the blueshifted gas for regions equidistant from the S nucleus.

A sample of the diverse H$_2$ S(1) line profiles in our data are shown in Figure \ref{fig:Vel_maps}, along with the best fit to the main and secondary components, where the main component is defined as the emission profile closest to the systemic velocity. We also plot the velocity maps of the main and secondary H$_2$ S(1) components in central panels of Figure \ref{fig:Vel_maps}. In the following two subsections, we describe the flux distribution and kinematics of the main and secondary components. 

\subsubsection{Main H$_2$ Gas Component} \label{subsec:main_comp}

The warm H$_2$ gas structure of the main component in the northern galaxy appears largely undisturbed while the gas in the southern galaxy appears asymmetric and complex. Specifically, as shown in the first column, first and third rows of Figure \ref{fig:Flux_maps}, there is a horizontal feature, likely the southern galactic disk seen edge on, that cuts across the S nucleus \citep{English2003}. There is also a vertical structure that reaches a peak luminosity roughly 350 pc south of the S nucleus. In terms of the kinematics, the redshifted emission seen in the northeast extends to the south, where it reaches a maximum velocity of +130 km s$^{-1}$ about 300 pc east of the S nucleus (near extraction region C in Figure \ref{fig:Vel_maps}). A similar structure of blueshifted gas to the west of the S nucleus is noticeably absent. Lastly, the velocity dispersion also shows a complex morphology throughout the FOV, and is largest immediately to the north and south of the S nucleus in the same regions where the secondary component is detected. In contrast, it is comparatively low in the regions surrounding the N nucleus.

\subsubsection{Secondary H$_2$ Gas Component} \label{subsec:outflow_comp}

As shown in Figure \ref{fig:Vel_maps} and the second and fourth rows of Figure \ref{fig:Flux_maps}, we detect secondary H$_2$ components that are collimated and symmetric around the S nucleus in the north-south direction. These secondary components are spectrally resolved, and we define their velocity as the offset from the systemic velocity of the S nucleus, as calculated from the 2.3 $\mu$m CO absorption bands. Based on our \textit{F}-test detection threshold, they are detected out to a projected distance of 650 and 500 pc to the north and south, respectively, beyond which their emission is too low to satisfy the \textit{F}-test. The northward, receding cone grows wider with increasing distance from the S nucleus (closer to the N nucleus), although it appears offset from the S nucleus and there is a noticeable shift eastwards. The spatial structure of southwards, approaching gas appears truncated, particularly in the S(1) emission, however this is likely due to the secondary component falling below our detection threshold (see Section \ref{sec:Analysis}). In both the approaching and receding gas, as seen in the velocity maps of Figure \ref{fig:Flux_maps}, the structure of the secondary component is generally cospatial with the cold CO gas seen in ALMA observations \citep{Sakamoto2014}. At shorter wavelengths (H$_2$ S(7) 5.5 $\mu$m), the secondary component is detected closer to the S nucleus, to about 80 pc, which can be seen in the second row of Figure \ref{fig:Flux_maps}.

The kinematics of the secondary component show maximum projected velocities of +340 km s$^{-1}$ in the redshifted gas and $-$310 km s$^{-1}$ in the blueshifted gas. The velocities appear relatively constant out to the edges of the detected emission, where the emission falls below our detection threshold about 600 pc away from the nucleus. The velocity dispersion is greatest, up to 140 km s$^{-1}$, 400 pc south of the S nucleus, and is generally constant ($\sigma\rm{_{avg}}$\;$\sim$\;90 km s$^{-1}$) elsewhere. 

To compare this to the warmer gas phase, which can be traced by H$_2$ 2.12 $\mu$m emission, we refer to VLT/SINFONI data which have a larger FOV than \textit{JWST}/NIRSpec. Here, peak projected velocities of +270 km s$^{-1}$ and $-$240 km s$^{-1}$, and a peak velocity dispersion of 190 km s$^{-1}$ are measured in H$_2$ 2.12 $\mu$m emission \citep{Emonts2014}, both of which are lower than the MIR component. The locations of these peaks, however, are cospatial.

\begin{deluxetable}{cccc}
\setlength{\tabcolsep}{8pt}
\caption{H$_2$ 0--0 Secondary Component Fluxes} 
\label{tab:H2_Fluxes}
\tablehead{\colhead{H$_2$ Line} & \colhead{Flux$\rm{_{Blue}}$} & \colhead{Flux$\rm{_{Red,N}}$} & \colhead{Flux$\rm{_{Red,S}}$}\\
\colhead{($\mu$m)} & \multicolumn{3}{c}{(10$^{-18}$ W m$^{-2}$)}\\
\colhead{(1)} & \colhead{(2)} & \colhead{(3)} & \colhead{(4)}}
\startdata
S(8) 5.05 & 1.30$\pm$0.10 & 0.18$\pm$0.02 & ---\\
S(7) 5.51 & 6.97$\pm$0.59 & 3.07$\pm$0.24 & 1.70$\pm$0.14\\
S(6) 6.11 & 2.56$\pm$0.21 & 0.74$\pm$0.06 & ---\\
S(5) 6.91 & 14.13$\pm$1.07 & 6.51$\pm$0.49 & 4.19$\pm$0.31\\
S(4) 8.02 & 4.31$\pm$0.34 & 3.34$\pm$0.25 & 1.67$\pm$0.13\\
S(3) 9.66 & 8.88$\pm$0.68 & 6.64$\pm$0.50 & ---\\
S(2) 12.28 & 3.75$\pm$0.31 & 4.68$\pm$0.41 & 2.12$\pm$0.19\\
S(1) 17.03 & 5.72$\pm$0.46 & 8.46$\pm$0.74 & 3.07$\pm$0.27\\
\hline \\[-1.8ex]
T$\rm{_{low}}$ (K) & 263$\pm$19 & 275$\pm$12 & 313$\pm$22\\
n & 4.54$\pm$0.11 & 5.89$\pm$0.13 & 5.32$\pm$0.22\\
M$\rm{_{out}}$ (10$^5\;M_{\odot}$) & 3.5$\pm$0.7 & 5.4$\pm$0.9 & 1.5$\pm$0.5
\enddata
\tablecomments{Columns: (1) H$_2$ emission lines with rest-frame wavelengths. (2) Flux within an aperture of radius $0\farcs8$ of the blueshifted secondary component, centered on its peak emission. The listed errors are representative of uncertainties in the fitting process. (3--4) Same as column 2, but for the redshifted secondary component within the northern and southern regions.\\
---Bottom rows: The final three rows list the lower gas temperature, power-law index, and mass of the secondary component gas for each region.}
\end{deluxetable}
\vspace{-4mm}

\subsection{Temperatures and Masses of the Secondary H$_2$ Gas Component} \label{subsec:Non-rotational_Mass}

The extensive wavelength coverage of \textit{JWST}, combined with its unprecedented sensitivity, allows us to obtain robust gas masses and temperatures of the warm H$_2$ gas directly from the rotational H$_2$ lines between 5--17 $\mu$m. Here, we calculate the mass of the secondary H$_2$ gas component using fluxes and power-law temperature modeling of the different transitions of H$_2$ \citep{Togi2016}. Aperture-corrected H$_2$ S(8)--S(1) fluxes were obtained using the r = $0\farcs8$ apertures centered on the peak of the blueshifted and redshifted secondary components (see the blue and red dashed circles in Figure \ref{fig:Flux_maps} and Table \ref{tab:H2_Fluxes}). This aperture size was chosen to contain as much of the emission as possible, while staying within the FOV of all MIRI channels. The third aperture was used to sample the redshifted gas closer to the S nucleus (see orange dotted circle in Figure \ref{fig:Flux_maps}). Its position is at a similar distance to the S nucleus as the aperture sampling the blueshifted gas so as to allow for a direct comparison of outflowing gas conditions. However, secondary, redshifted H$_2$ S(3), S(6) and S(8) emission were not included in the fit due to not passing our S/N $>$ 5 detection threshold. Extinction correction was done using the 9.7 $\mu$m silicate absorption, where corrections were generally less than 10$\%$ and were slightly higher in the blueshifted gas. Only the H$_2$ S(3) line at 9.66 $\mu$m had relatively large corrections, up to 25$\%$.

\begin{figure}
\epsscale{1.2}
\plotone{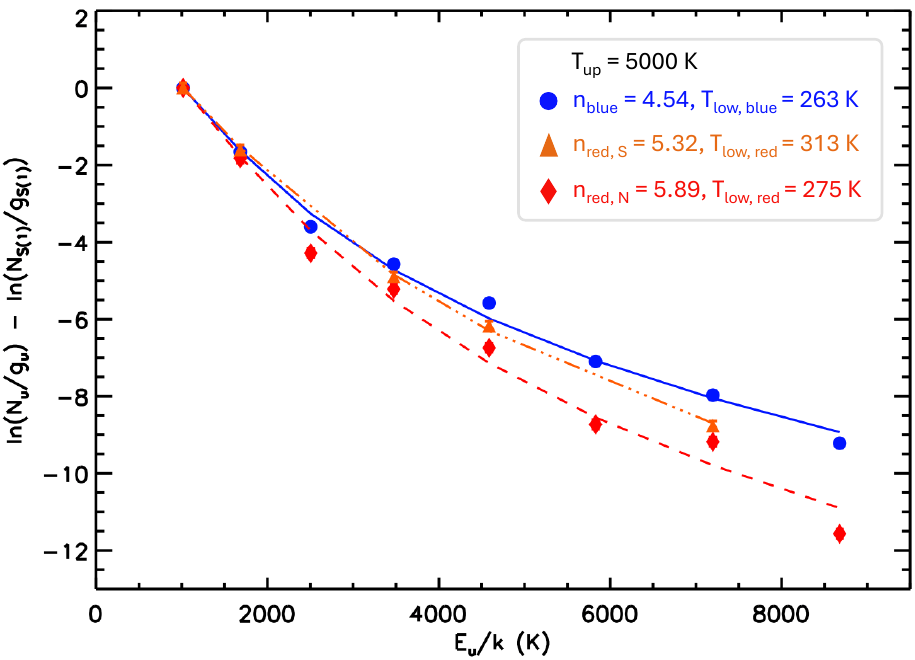}
\caption{Excitation diagram of the blue and redshifted gas using the r=$0\farcs8$ apertures, as shown in Figure \ref{fig:Flux_maps}. Plotted are the column densities ($N$) of the H$_2$ S(8)--S(1) transitions, normalized by their statistical weight, $g$, as a function of their energy level $E$. The $N/g$ ratios are normalized with respect to the S(1) transition. The best fits to the redshifted (northern and southern regions) and blueshifted gas are denoted by the dashed red, orange, and solid blue lines, respectively. H$_2$ S(3), S(6) and S(8) emission were not included in the southern portion of the redshifted gas due to not passing our S/N $>$ 5 detection threshold. The errors are comparable to the symbol size. The lower power-law index of the blueshifted gas indicates that it has a larger warm gas mass fraction. In addition, the fraction of warm gas seems to be higher closer to the nucleus. \label{fig:excitation}}
\end{figure}

To obtain mass and temperature estimates, we use the power-law model described in \cite{Togi2016}. The power-law index, lower, and upper gas temperatures are the primary model parameters. Since varying the temperature when it is greater than 5000 K has a negligible effect on the quality of the fit and the gas mass, we fixed the upper temperature to 5000 K while the other two parameters were treated as free variables. The model assumes local thermodynamic equilibrium and samples a range of power-law indices and lower gas temperatures, the latter of which determines the ortho-to-para ratio based on Figure 1 of \cite{Sternberg1999}. H$_2$ line fluxes were used to estimate column densities, whose ratios were then used to create an excitation diagram (see Figure \ref{fig:excitation}). A best-fit model was then fit to the column densities, from which the lower temperature and power-law index were used to calculate the total column density and gas mass.

Fits to the secondary H$_2$ gas component result in a lower gas temperature of 263$\pm$19 K, 275$\pm$12 K, and 313$\pm$22 K (ortho-to-para ratio of $\sim$3), and a power-law index of 4.54$\pm$0.11, 5.89$\pm$0.13, and 5.32$\pm$0.22 for the blueshifted, redshifted (north), and redshifted (south) gas, respectively. In other local LIRGs, typical power law indices range from 3.4 to 5.4 based on \textit{Spitzer/IRS} data \citep{Togi2016}, which are consistent with our values. Our fits result in a warm molecular gas mass (including a heavy element correction factor of 1.36) of $M\rm{_{blue}}$ = (3.5$\pm$0.7)$\times$10$^5\;M_{\odot}$, $M\rm{_{red,north}}$ = (5.4$\pm$0.9)$\times$10$^5\;M_{\odot}$, and $M\rm{_{red,south}}$ = (1.5$\pm$0.5)$\times$10$^5\;M_{\odot}$. Since these apertures do not fully encompass the full extent of the secondary component, the masses calculated here do not represent the total gas mass associated with the secondary component.

To account for the gas mass missed by our apertures, we assumed the power-law indices as calculated from the $0\farcs8$ extractions are constant throughout the secondary component gas. Since channel 3 has the largest FOV in our analysis and is thus likely to best trace the full extent of the secondary component, we opted to use the two H$_2$ lines, S(1) and S(2), that are present in this channel. For both S(1) and S(2), we found that the $0\farcs8$ apertures at the emission peak encompasses about 60$\%$ of the total detected flux in the blueshifted gas. Assuming a constant power-law index of 4.54 and a temperature of 263 K, this implies that the molecular mass could be as high as $M\rm{_{blue}}$ = (5.8$\pm$1.2)$\times$10$^5\;M_{\odot}$. For the redshifted gas, we found that the northern and southern apertures encompass roughly 65$\%$ and 20$\%$ of the total detected redshifted flux in both S(1) and S(2). Extrapolating using the northern region values, n=5.89 and T$\rm{_{low}}$=275 K, leads to a total estimated gas mass of $M\rm{_{red}}$ = (8.3$\pm$1.4)$\times$10$^5\;M_{\odot}$. Similarly, extrapolating using the southern region values, n=5.32 and T$\rm{_{low}}$=313 K, results in a total gas mass of $M\rm{_{red}}$ = (7.9$\pm$2.6)$\times$10$^5\;M_{\odot}$. These two estimates are in good agreement, and we opt to use (8.3$\pm$1.4)$\times$10$^5\;M_{\odot}$ since the aperture used encompasses a larger fraction of the total emission.

However, the above calculations only include the S(8)--S(1) transitions, and not the H$_2$ S(0) 28.22 $\mu$m line, which falls outside the wavelength coverage of our MIRI/MRS data. The inclusion of this line tends to lower the estimated temperature, which, in turn, would result in an increase of the calculated gas mass. While the masses listed above should be considered lower limits, we use $M\rm{_{total}}$ = (5.8$\pm$1.2)$\times$10$^5\;M_{\odot}$ + (8.3$\pm$1.4)$\times$10$^5\;M_{\odot}$ = (1.4$\pm$0.2)$\times$10$^6\;M_{\odot}$ as the total gas mass of the secondary component for our analysis.

In addition to large aperture extractions, we can also investigate the temperature distribution of the secondary component on a spaxel-by-spaxel basis. Here, we opt to use the S(1) and S(7) transitions, which are excited at low and high temperatures, respectively. Although S(8) is excited at higher temperatures than S(7), its lower S/N makes it less suited for this analysis.

To compare the S(1) and S(7) maps on the same angular resolution, we first convolved the S(7) flux map (see second row, first panel of Figure \ref{fig:Flux_maps}). The convolution was performed using a Gaussian kernel, which reduced the resolution of the S(7) map (FWHM=$0\farcs25$) to that of the S(1) map (FHWM=$0\farcs7$). To account for the difference in pixel size of the two maps, we subsequently rebinned the convolved S(7) map to the larger pixel size of the S(1) map. We then calculated the S(7)/S(1) ratios by dividing the two maps, where pixels with S/N $>$ 5 in both maps were kept (see Section \ref{sec:Analysis}).

\begin{figure}
\epsscale{0.9}
\centering
\plotone{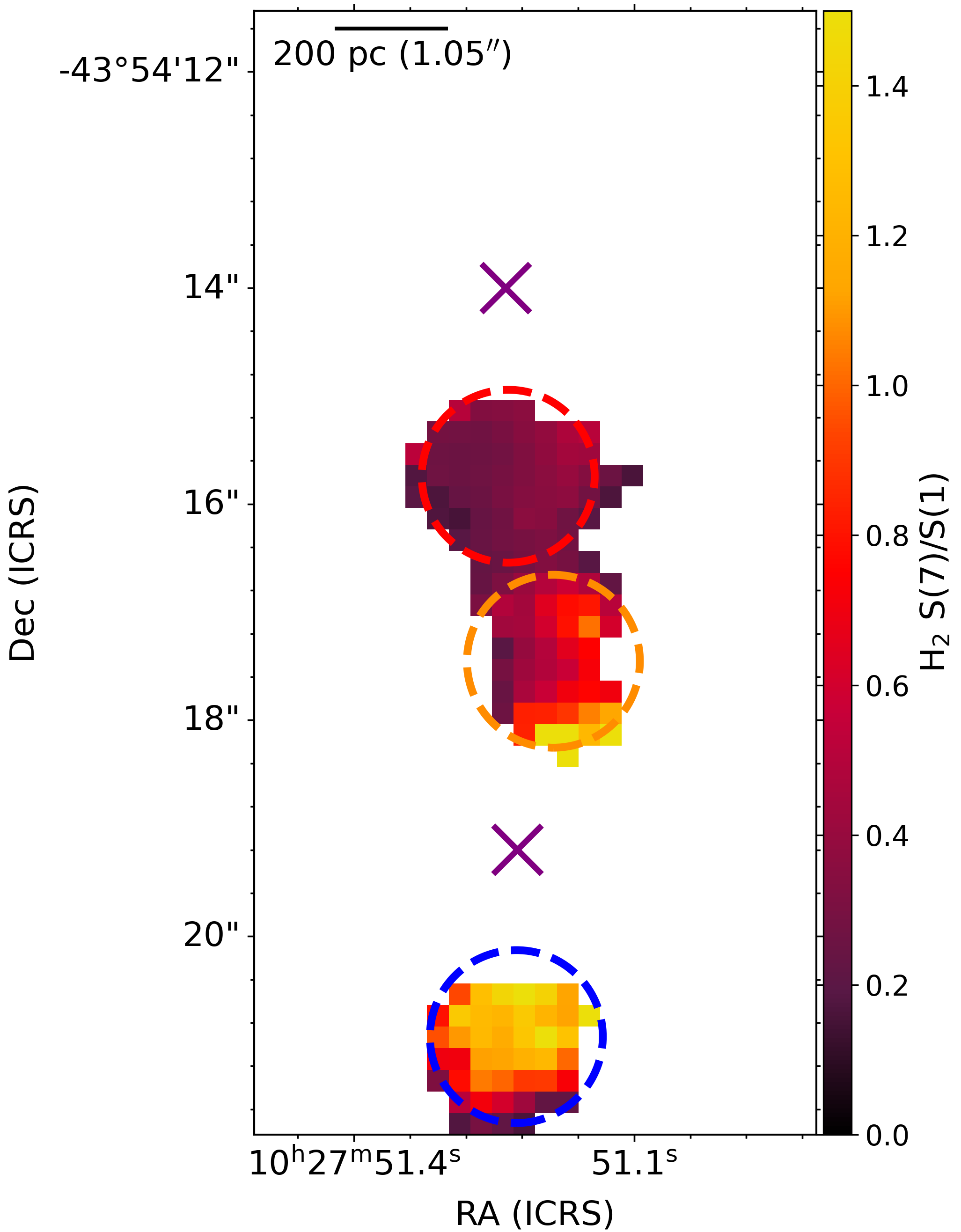}
\caption{H$_2$ S(7)/S(1) flux ratio map. The purple crosses indicate the locations of the two nuclei, and the apertures used to calculate the masses and power-law indices of the secondary component are shown as dashed circles. We see higher ratios, indicating a larger fraction of warmer gas, closer to S nucleus, suggesting that the heating of the H$_2$ is caused by it. We also measure slightly higher average ratios in the southwards, blueshifted gas than the northwards, redshifted gas. Lastly, the temperature distribution within the apertures are indicative of a larger warm gas fraction in blueshifted gas.
\label{fig:H2_ratio}}
\end{figure}

The resulting ratio map is shown in Figure \ref{fig:H2_ratio}, where higher line ratios indicate a larger warm gas mass fraction. We find the largest warm gas fractions close to the S nucleus, with the ratios decreasing at larger radii. Our calculated power-law indices for the blueshifted (4.54) and redshifted (5.32 and 5.89) outflows suggest a higher warm gas fraction in the blueshifted outflow, which is also evident from the temperature distribution. In the redshifted, northwards gas, we measure S(7)/S(1) ratios of $\sim$0.6 at a distance of 300 pc from the S nucleus. This decreases to $\sim$0.3 at 600 pc, where the H$_2$ fluxes reach their peak in the redshifted gas. For a given distance from the S nucleus, we find slightly higher average ratios in the blueshifted gas, 1.3 at 300 pc, compared to the redshifted gas, 0.6. Unfortunately, we are not able to assess the degree to which the temperature changes beyond 450 pc from the S nucleus, due to the secondary component expanding beyond the MIRI channel 1 FOV.

\subsection{Gas Excitation of the Interstellar Medium} \label{subsec:Excitation_analysis}

The transitions of H$_2$, along with ionized hydrogen recombination lines and [\ion{Fe}{2}], can be used to analyze the gas excitation of the ISM. H$_2$ can be excited through thermal processes, such as collisional excitation, as well as nonthermal processes, such as UV photon absorption. In each case, flux ratios of H$_2$ lines can be used to help distinguish the different mechanisms \citep{Guillard2009,Mazzaalay2013,Petric2018,U2019}. Iron can also serve as a tracer of shocks, since it is mostly embedded in dust grains and released only when these grains are destroyed through thermal sputtering caused by shocks \citep{Storchi2009,Riffel2013,Inami2013}. Due to the relative abundance of these lines, trends between H$_2$/H and [\ion{Fe}{2}]/H have been used to help distinguish excitation sources \citep{Larkin1998,Riffel2013,Colina2015,U2022a,Bianchin2024}.

\begin{figure*}
\gridline{\fig{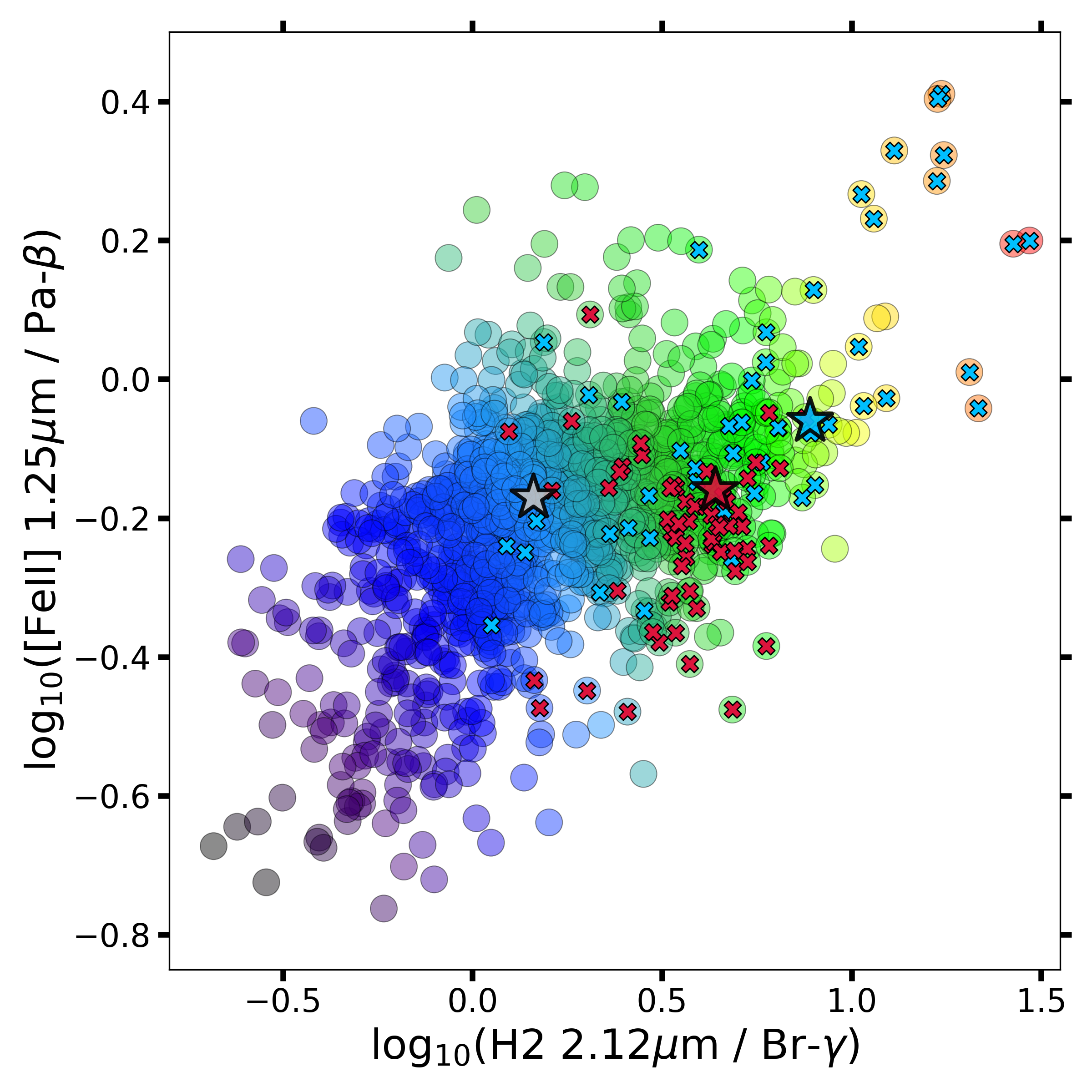}{0.43\textwidth}{}
\fig{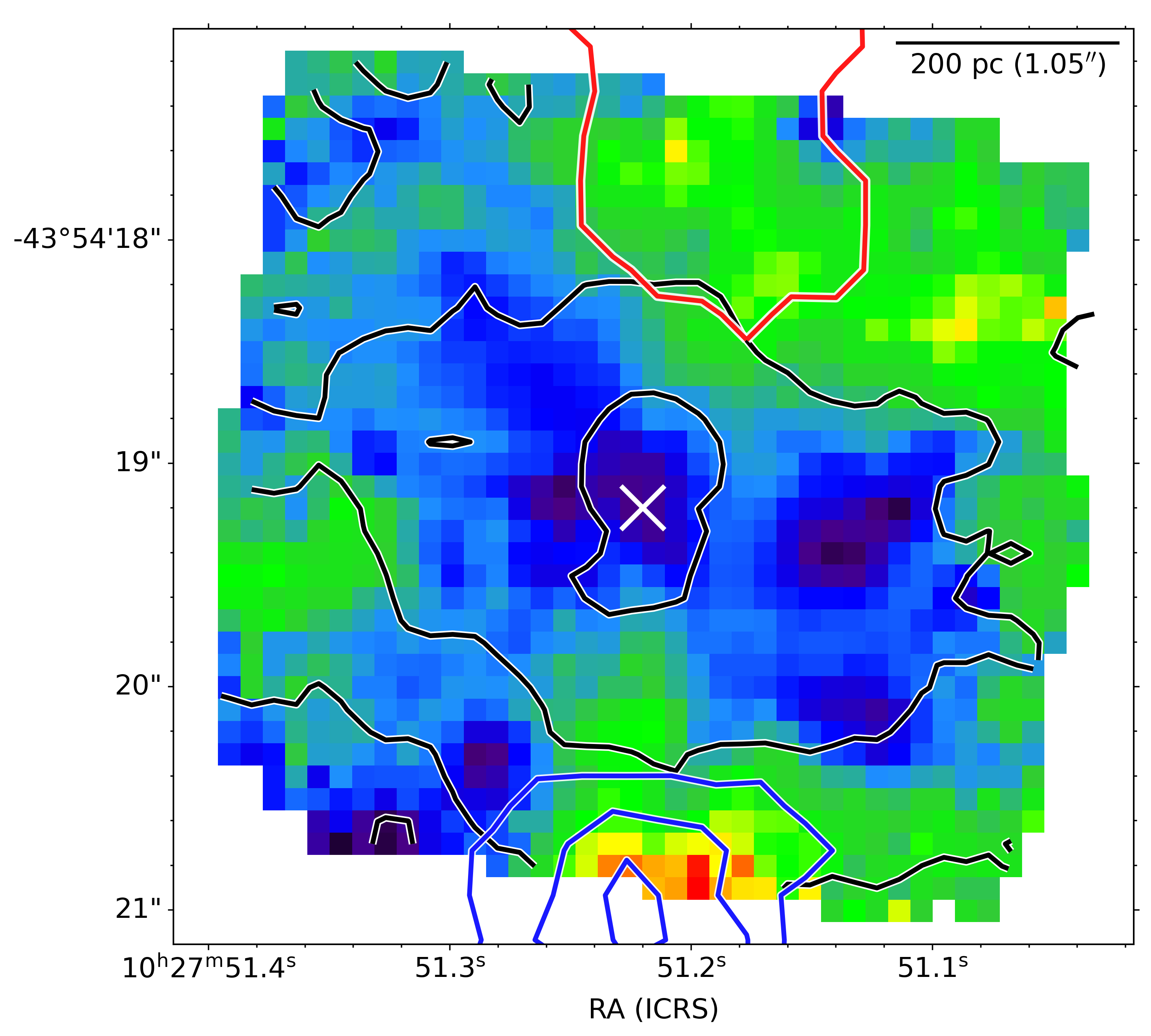}{0.48\textwidth}{}}
\caption{[\ion{Fe}{2}] 1.25 $\mu$m/Pa$\beta$ versus H$_2$ 1-0 S(1) 2.12 $\mu$m/Br$\gamma$ diagram (\textit{left}) and associated excitation map of the S nucleus (\textit{right}). Each spaxel in the full NIRSpec FOV on the right is color coded based on a linear fit to the line ratios, as shown in the diagram in the left-hand panel. Data points filled with crosses correspond to spaxels found within the outflow regions, where the cross color identifies whether it is the blue or redshifted outflow. Median ratio values of the regions within the blue outflow, red outflow, and regions exterior to the outflow are shown as blue, red, and gray stars, respectively. Blue and red contours represent the secondary component (outflows) as traced by the H$_2$ S(1) 17.0 $\mu$m line (see Figure \ref{fig:Flux_maps}), while the black contour represents the PAH 3.3 $\mu$m emission, set at 1.0, 5.0, and 10.0$\times$10$^{-19}$ W m$^{-2}$. The white cross at the center marks the location of the S nucleus. The regions where we detect the secondary H$_2$ components (outflows) tend to have higher [\ion{Fe}{2}]/Pa$\beta$ and H$_2$/Br$\gamma$ ratios that are not seen in the regions surrounding the S nucleus. \label{fig:Excitation_map}}
\end{figure*}

In the left panel of Figure \ref{fig:Excitation_map}, we explore the spatial variations of gas excitation by investigating the relation between the ratios [\ion{Fe}{2}] 1.25 $\mu$m/Pa$\beta$ versus H$_2$ 1-0 S(1) 2.12 $\mu$m/Br$\gamma$ in the nuclear region of the southern galaxy. The purpose of this relation is to identify enhanced emission which could be caused by shocks, and we thus use the total emission of each line. To show the spatial distribution of the line ratios, we color code every spaxel in accordance to a linear fit to the ratios, producing an excitation map covering the full NIRSpec FOV, as shown in the right panel of Figure \ref{fig:Excitation_map}. The excitation structure is complex, but there is a noticeable peak in the line ratios to the south of the S nucleus. The median values of the ratios in this region are [\ion{Fe}{2}]/Pa$\beta$ = -0.06 and H$_2$/Br$\gamma$ = 0.89, compared to -0.17 and 0.16 for regions outside the outflow (see blue and gray stars in left panel of Figure \ref{fig:Excitation_map}). To the north of the nucleus, there is a broad structure with slightly higher excitation, median ratios of [\ion{Fe}{2}]/Pa$\beta$ = -0.16 and H$_2$/Br$\gamma$ = 0.64 (red star in Figure \ref{fig:Excitation_map} left), that extends westwards. The lowest line ratios are located in the S nucleus and in regions coincident with multiple clumps that are detected in NIRCam imaging \citep{Linden2024}. The implications of these regions and corresponding line ratios are discussed in Section \ref{subsec:shock_discussion}.

\section{Discussion} \label{sec:Discussion}

The high velocities and structure of the secondary H$_2$ emission-line component are most consistent with outflowing gas, and we hereafter refer to the secondary component as the outflow component. Although outflowing warm H$_2$ gas has previously been identified \cite[e.g.,][]{Emonts2017,U2019}, this is the first time that the warm, outflowing H$_2$ gas in the MIR range is spatially and spectrally resolved. 

It is noteworthy, however, that we do not detect a secondary H$_2$ component in or immediately around the N nucleus. Specifically, our MIR spectra do not show H$_2$ lines with sufficient asymmetry to pass our detection thresholds (see Section \ref{sec:Analysis}) in the regions where the cold molecular outflow from the N nucleus is detected \citep{Sakamoto2014,Michiyama2018}. As such, for the remainder of this article, we will only reference and analyze the outflows associated with the S nucleus.

\subsection{Outflow Characteristics} \label{subsec:Outflow_kinematics}

The symmetry of the outflow structure to the north/south of the S nucleus indicates that it is likely driven by the S nucleus. However, as shown in the spatial maps of Figure \ref{fig:Flux_maps}, we do not detect outflow emission within 80 pc of the nucleus. This is likely because the emission either does not satisfy our S/N $>$ 5 detection threshold, or has a velocity that is too low to be separated from the main gas component. The latter is supported by the higher velocity dispersion in the regions immediately surrounding the nucleus (see right panels of Figure \ref{fig:Flux_maps}), where a spectrally unresolved component could broaden the line profile of the main component, thus making detection of a secondary component more difficult. Additionally, coupled with the fact that the spectral resolution of MIRI is lower for S(1), lower outflow velocities that cannot be spectrally resolved could be a reason why we do not detect S(1) emission as close to the nucleus as S(7). Close inspection of the S(1) line profiles do show faint wings, possibly part of an unresolved component, at similar proximities to the S nucleus as S(7). However, adding a secondary component only marginally improved the fit and did not pass our \textit{F}-test of variances used for outflow detections. While it is likely that S(1) emission is present closer to the S nucleus, it is difficult to perform robust fits and characterize the outflow parameters within 80 pc.

The outflows are detected out to a distance of 650 and 500 pc to the north and south of the nucleus. To obtain deprojected distances, we adopt the outflow inclination as calculated by \cite{Sakamoto2014} and \cite{Emonts2014}, $i_{\rm{out}}\sim20^\circ$, which results in intrinsic distances of 690 and 530 pc. The regions with the highest outflow emission are spatially coincident with strong emission in the main H$_2$ component and cold CO gas. Moreover, the distribution of the HCN and HCO$^+$ gas, which trace the denser gas, show enhanced emission in these outflow regions \citep{Michiyama2018}. As such, these H$_2$ flux peaks could be caused by the outflowing gas impacting a dense region in the ISM. 

The kinematics of the outflows show maximum projected velocities to be +340 km s$^{-1}$ to the north and $-$310 km s$^{-1}$ to the south. Deprojecting the outflows, v$\rm{_{out}}$/sin$(i_{\rm{out}})$, we obtain maximum intrinsic velocities of +990 and $-$910 km s$^{-1}$, respectively (where v$\rm{_{out,avg}}$= 730 km s$^{-1}$, averaged across all regions of the outflows). 

Inspection of the outflow spatial maps (Figure \ref{fig:Flux_maps}) shows that the northwards (redshifted) outflow appears to have a marginal eastward shift as it expands outwards. Based on the relative location of the two galaxies and the interlaced nature of their disks (see NIRCam imaging in Figure \ref{fig:RGB_plot}), this shift eastwards could be induced by the disk rotation of the northern galaxy. If the northern disk is rotating clockwise \citep{Sakamoto2014}, then this could potentially shift the outflowing gas eastward, explaining the angled structure relative to the S nucleus. 

\subsection{Outflow Mass Comparisons of MIR to NIR and Submillimeter Measurements} \label{subsec:Mass_compare}

In Section \ref{subsec:Non-rotational_Mass}, we calculated the total mass of the warm outflowing gas to be $M\rm{_{blue,out}}$ = (5.8$\pm$1.2)$\times$10$^5\;M_{\odot}$ and $M\rm{_{red,out}}$ = (8.3$\pm$1.4)$\times$10$^5\;M_{\odot}$ for the blueshifted and redshifted outflows, respectively. We can compare these masses to that of the hotter ($\sim$2000 K) gas observed at NIR wavelengths. To measure this hotter component, we use the H$_2$ 1-0 S(1) 2.12 $\mu$m outflow flux map created by \cite{Emonts2014}, which was derived from K-band VLT/SINFONI observations (program ID: 078.B-0066(A), PI: Luis Colina). Measuring the flux within the same apertures used to calculate the warm MIR gas mass and following the methods described in \cite{Mazzaalay2013} and \cite{Emonts2014}, we obtain mass measurements of 330$\pm$60 and 200$\pm$35 $M_{\odot}$ for the blue and redshifted outflows, respectively. This combined gas mass of 530 $M_{\odot}$ is over 2.5$\times$10$^3$ times lower than the warm MIR component, (1.4$\pm$0.3)$\times$10$^6\;M_{\odot}$. 

In order to assess the contribution of the warm gas to the total multiphase component, we can use molecular CO to estimate the mass of the outflowing cold H$_2$ gas. To do this, we opted to use archival CO (2-1) ALMA data (project code: 2015.1.00902.S, PI: Yiping Ao), which has higher spatial resolution than the CO (1-0) map\footnote{The spatial resolution of the CO (2-1) data is $1\farcs03\times0\farcs56$ while the CO (1-0) data is $1\farcs6\times1\farcs2$.} of \cite{Sakamoto2014}. Using the same two r=$0\farcs8$ apertures for our H$_2$ mass calculations, we extracted spectra of the CO (2-1) emission. Based on a two-component Gaussian fit, we measured line fluxes of 20.6$\pm$2.1 and 8.6$\pm$1.9 Jy km s$^{-1}$ with velocity offsets of $-$150 km s$^{-1}$ and $+$250 km s$^{-1}$ for the blueshifted and redshifted gas, respectively. These are consistent with the mean velocities measured in the warm H$_2$ outflow, $-$170 km s$^{-1}$ and $+$240 km s$^{-1}$. Following \cite{Michiyama2018}, we set L$\rm{_{CO(2-1)}}$/L$\rm{_{CO(1-0)}}$ as unity. For the CO to H$_2$ conversion factor, $\alpha\rm{_{CO}}$=$M\rm{_{mol}}$$/L\rm{_{CO}}$, we assume a value of 0.8 $M_\odot$ (K km s$^{-1}$ pc$^2$)$^{-1}$, which is typically used for ULIRGs \citep{Downes1998,Bryant1999}. This value is 5.4 times lower than the galactic value of 4.32. This is because ULIRGs, particularly those involved in mergers, typically have increased CO line widths (and thus greater L$\rm{_{CO}}$) caused by the heightened turbulence and temperatures induced by the outflows and mergers \citep{Togi2016}. From these assumptions, we obtained $M\rm{_{blue,out}}$ = (1.6$\pm$0.2)$\times$10$^7\;M_{\odot}$ and $M\rm{_{red,out}}$ = (6.8$\pm$1.5)$\times$10$^6\;M_{\odot}$, which implies the combined outflowing mass is $M\rm{_{cold,out}}$ = (2.3$\pm$0.2)$\times$10$^7\;M_{\odot}$. 
 
While we assumed a value of 0.8 for $\alpha\rm{_{CO}}$, \cite{Cicone2018} calculated $\alpha\rm{_{CO}}$ to be 2.1 where molecular outflows are present. This in turn would increase our cold gas mass estimate to $M\rm{_{cold,out}}$ = (6.0$\pm$0.8)$\times$10$^7\;M_{\odot}$. More recently, \cite{Pereira2024} estimated $\alpha\rm{_{CO}}$ of various regions around the outflows in NGC 3256 utilizing the CO rovibrational band at 4.67 $\mu$m. From their two-component model, they report $\alpha\rm{_{CO}}$ to be 0.62. Setting $\alpha\rm{_{CO}}$ as 0.62 reduces our combined cold gas mass to (1.8$\pm$0.3)$\times$10$^7\;M_{\odot}$, still consistent with the mass calculated using the canonical conversion factor for local ULIRGs. Hereafter, we adopt 2.3$\times$10$^7\;M_{\odot}$ for the remainder of our analysis.



Combining the redshifted and blueshifted outflows, the $M\rm{_{warm}}$/$M\rm{_{cold}}$ gas mass fraction is about 6$\%$. As mentioned previously, we note that our data set does not include the S(0) 28.22 $\mu$m line which tends to lower the estimated temperature, thus increasing the $M\rm{_{warm}}$/$M\rm{_{cold}}$ fraction. \cite{Higdon2006} measured typical warm gas mass fractions in observed ULIRGs to be slightly less than 1$\%$ (as calculated from the S(1) 17.0 $\mu$m to S(3) 9.66 $\mu$m line fluxes). However, they found that the inclusion of S(0) could increase $M\rm{_{warm}}$/$M\rm{_{cold}}$ to tens of percent. While the warm mass fraction may depend heavily on the inclusion of S(0), our estimates for the warm mass fraction in the outflows are comparable to those measured in other local active galaxies, which generally range from 1--10$\%$, though it can reach up to 35$\%$ in some cases \citep{Rigopoulou2002}.


In Section \ref{subsec:Outflow_kinematics}, we calculated the maximum intrinsic outflow velocity to be $\sim$10$^3$ km s$^{-1}$, and maximum deprojected distance to be 700 pc. Following this, the outflow time scale, $t\rm{_{out}}$, is estimated to be $7\times10^5$ yr. Assuming a conical geometry, this results in a time-averaged outflow mass rate of $\dot{M}\rm{_{warm,out}}$ = 2.0$\pm$0.8 M$_{\odot}$ yr$^{-1}$. If a significant portion of the outflow mass is in the cold component, then the outflow mass rate could be as high as $\dot{M}\rm{_{cold,out}}$ = 33 M$_{\odot}$ yr$^{-1}$, which is slightly lower than the 48 M$_{\odot}$ yr$^{-1}$ reported by \cite{Sakamoto2014}, though our value here should be considered a lower limit due to our use of a r = $0\farcs8$ aperture. Regardless, these values are comparable to rates seen in other Seyferts and ULIRGs \citep{Garcia2011,Cicone2014}. These rates, particularly the cold gas mass rate, are high compared to estimates of the SFR. In regions around the S nucleus, \cite{Emonts2014} and \cite{Michiyama2018} estimated the SFR to be 1 - 6.8 M$_{\odot}$ yr$^{-1}$ based on hydrogen recombination line calibrations. For the cold molecular outflow, this leads to a mass loading factor, $\dot{M}\rm{_{cold,out}}$/SFR, of five or greater. Amongst starburst galaxies, mass loading factors are typically around unity while AGN-dominated sources tend to be higher ($>3$) \citep{Cicone2014}.

\subsection{Gas Conditions in the Outflow} \label{subsec:shock_discussion}

To investigate the impact the outflows have on the local ISM, we plot [\ion{Fe}{2}]/Pa$\beta$ versus H$_2$/Br$\gamma$ line ratios in Figure \ref{fig:Excitation_map}. Higher [\ion{Fe}{2}] and H$_2$ ratios relative to the recombination lines indicate stronger gas excitation, and are located at the southern edge of the FoV, where they are spatially coincident with the blueshifted outflow. This region is also where the outflow flux is the strongest (see Figure \ref{fig:Flux_maps}) and has the highest fraction of warm gas (see Figure \ref{fig:H2_ratio}). This increase in the H$_2$ and [\ion{Fe}{2}] ratios could be caused by the outflow impacting a denser region in the ISM, possibly in an interaction with the disk of the northern galaxy. This interpretation can be supported by the comparatively low value of the power-law index of the southern outflow, 4.54, which is lower than the mean found in other galaxies, 4.84 \citep{Togi2016}. Indices in this range are seen in other cases where shocks are present \citep{Appleton2017}, suggesting a shock front created by the outflow. However, most of the outflow falls outside the NIRSpec FOV, so it is difficult to assess the full effect of the excitation using this method. While the narrow FOV limits us from analyzing the redshifted outflow as well, the outflow regions that do fall within the NIRSpec FOV of the north galaxy also show elevated values for these line ratios, indicating that the northwards outflow may also be energizing the ISM.


\subsection{Outflow Energetics and Feedback} \label{subsec:Outflow_energetics}

\subsubsection{Energy of the Outflow} \label{subsubsec:Outflow_energy}

To explore the energy source of the outflows, we can calculate the combined kinetic energy of the outflows by combining turbulent and bulk components, 

\begin{equation}
\label{eq:outflow_vel}
E{\rm{_{kin}}} = E{\rm{_{turb}}} + E{\rm{_{bulk}}} =  \frac{3}{2}M\sigma^2 + \frac{1}{2}Mv\rm{_{out}}^2
\end{equation}

Setting $M=M\rm{_{warm,out}}=1.4\times10^6\;M_{\odot}$, $\sigma\rm{_{avg}}$\;$\sim$\;90 km s$^{-1}$, and $v\rm{_{out,avg}}\;\sim$\;730 km s$^{-1}$ gives a combined outflow energy of $E\rm{_{warm,kin}}\;\sim\;8\times10^{54}\;$ erg. For the cold H$_2$ outflow, where $M\rm{_{cold,out}}=2.3\times10^7\;M_{\odot}$, and assuming the same kinematics yields $E\rm{_{cold,kin}}\;\sim\;1\times10^{56}$ erg. Assuming the outflow time scale to be $7\times10^5$ yr, this results in a kinetic power of $P\rm{_{warm,out}}\;\sim\;4\times10^{41}$ erg s$^{-1}$ and $P\rm{_{cold,out}}\;\sim\;6\times10^{42}$ erg s$^{-1}$.

Put together, the temperatures, kinematics, and energetics of the outflows can provide clues to their origin. As shown in Figure \ref{fig:H2_ratio}, we find the highest warm gas fractions close to the S nucleus, with fractions decreasing at larger radii, an indication of heating from the nuclear source. Moreover, in addition to the symmetry to the north and south of the S nucleus, the fast velocities ($\sim$10$^3$ km s$^{-1}$), large mass loading factor ($>5$), and power ($6\times10^{42}$ erg s$^{-1}$) found here are all typical of AGN-driven outflows \citep{Rodriguez2006,Muller2011,Hill2014,Bae2016,Harrison2016,Veilleux2020}.

\subsubsection{Feedback on Star Formation} \label{subsubsec:SF_feedback}

It is well known that sustained outflows can suppress star formation \citep[e.g.,][]{King2015,Armus2020}. A commonly used IR tracer of star formation are PAHs \citep{Peeters2004,Inami2018,Lai2020,Lai2022}. Because PAH grains are fragile, they can easily be destroyed in harsh radiation environments. As such, we can utilize PAH fluxes and ratios to assess if the outflows have any significant feedback on the local ISM. 

\begin{figure*}
\centering
\epsscale{1.15}
\plotone{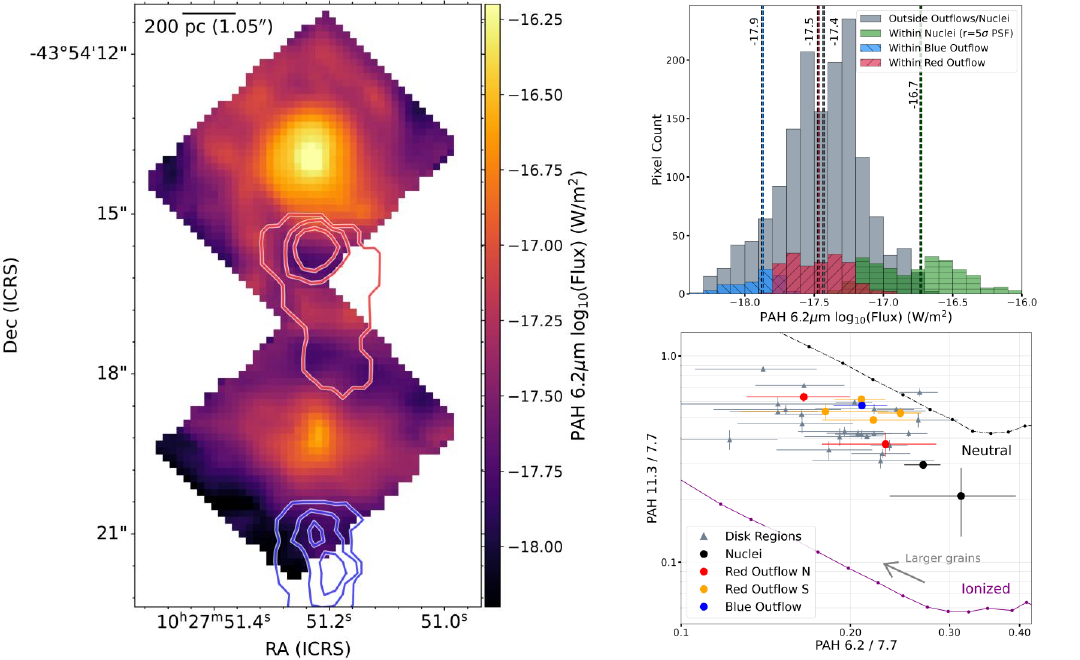}
\caption{(\textit{left}) PAH 6.2 $\mu$m flux map, plotted over the full MIRI channel 1 FoV. Blue and red contours mark the regions where we detect the molecular outflows, as traced by H$_2$ S(1) 17.0 $\mu$m emission. The PAH emission is strongest in the nuclear region of each galaxy, with prominent emission also seen in the disks and spiral arms. (\textit{upper-right}) PAH 6.2 $\mu$m flux histogram. Regions within each outflow (red and blue) and the nuclei (green) have been separately identified from those outside them (gray). The outflow regions are based on the contours in Figure \ref{fig:PAH}, and the nuclear regions are defined as the central region within a radius of 5$\sigma$ of the PSF. The dashed lines indicate the median value for each region and are accordingly labeled. (\textit{lower-right}) PAH$_{6.2}$/PAH$_{7.7}$ and PAH$_{11.3}$/PAH$_{7.7}$ flux ratios from \textsc{CRETA} grid extractions of the disk, nuclear, and outflow regions of the northern and southern galaxy (see the Appendix). Model tracks from \citep{Draine2021} show the theoretical values for neutral and ionized PAHs. The ratios of the outflow regions are within the scatter of the disk ratios, suggesting that the outflows have no significant effect on the grain size and ionization state of the PAHs. \label{fig:PAH}}
\end{figure*}

In the right panel of Figure \ref{fig:Excitation_map}, we plot 3.3 $\mu$m PAH flux contours over the excitation map. Most of the PAH flux is centered around the S nucleus with emission extending to the east and west, likely tracing the disk of the southern galaxy. The emission seems to decrease in the outflow regions but the limited FOV makes it difficult to ascertain the exact correlation between the outflow and PAH emission. As such, we can take advantage of a larger FOV of MIRI to examine the PAH at 6.2 $\mu$m, as shown in the flux map of Figure \ref{fig:PAH}. 

Like PAH 3.3 $\mu$m, we measure the strongest PAH emission at the nuclear region of each galaxy, with other prominent emission seen in the disks and spiral arms, likely caused by the star formation occurring there. As was done in Figure \ref{fig:Flux_maps}, we overlay the outflowing gas as traced by the H$_2$ S(1) 17.0 $\mu$m. Compared to the regions outside the outflows and nuclei, the median PAH flux in the regions coincident with the redshifted outflow is similar: log($F$/W m$^{-2}$) = -17.5 compared to -17.4 (see upper-right panel of Figure \ref{fig:PAH}). Fluxes within the blue outflow, however, have a lower median value: log($F$/W m$^{-2}$) = -17.9. Looking at the larger-scale structure of this merging system, we note that the northern galactic disk is roughly face on, $i_{inc,N}\sim30^\circ$, while the southern disk is edge on, $i_{inc,S}\sim80^\circ$ \citep{Sakamoto2014}. Based on this geometry and the location of the nuclei, the lower PAH fluxes within the blue outflow are likely due to the outflow not being co-spatial nor in projection with the inner, more dense regions of the northern disk or spiral arm structure where PAH emission is enhanced due to active star formation. The red outflow, on the other hand, is in projection and potentially co-spatial with these inner regions of the northern galaxy.

The PAH grain size and ionization state, which can be traced through the PAH$_{6.2}$/PAH$_{7.7}$ and PAH$_{11.3}$/PAH$_{7.7}$ ratios, can also provide insight to the state of the ISM and local star formation \citep{Rigopoulou2021,Lai2022}. To calculate these ratios in NGC3256, we use the CAFE Region Extraction Tool Automaton (\textsc{CRETA}\footnote{\url{https://github.com/GOALS-survey/CAFE/tree/master/CRETA}}, Diaz-Santos et al., \textit{in prep.}) to perform $1\farcs0\times1\farcs0$ grid extractions covering the full FoV of MIRI channel 1 (see Figure \ref{fig:PAH_ratio_map} in the Appendix). In order to robustly fit the 6.2 $\mu$m, 7.7 $\mu$m complex, and 11.3 $\mu$m complex PAH emission features with the continuum, we opted to use the Continuum And Feature Extraction (\textsc{CAFE}\footnote{\url{https://github.com/GOALS-survey/CAFE}}, Diaz-Santos et al., \textit{in prep.}) code. PAH flux ratios were calculated based on these fits and we plot these in the spatial maps of Figure \ref{fig:PAH}. These ratios are also listed in Table \ref{tab:PAH_ratios} of the Appendix.

In the lower-right panel of Figure \ref{fig:PAH}, we plot PAH$_{11.3}$/PAH$_{7.7}$ versus PAH$_{6.2}$/PAH$_{7.7}$. Model tracks of theoretical values for neutral and ionized PAHs from \cite{Draine2021} are also overplotted, where we have assumed the \cite{Bruzual2003} interstellar radiation field with a radiation strength parameter log\textit{U} = 2. Grids containing outflow emission have also been highlighted. The range of these ratios in the outflow regions (mean values of 0.21 and 0.54 for PAH$_{6.2}$/PAH$_{7.7}$ and PAH$_{11.3}$/PAH$_{7.7}$, respectively) are all within those of the disk values: 0.12 -- 0.27 and 0.30 -- 0.85. As such, although the outflows we detect are fast and significantly heat the molecular gas, they show no clear signs of altering the average PAH ionization state or size compared to other regions in the disk of NGC 3256, nor do they significantly reduce the surface brightness of the PAH emission in the regions of the wind. However, it is difficult to ascertain whether this lack of feedback is due to the outflow crossing over the galaxy disk (in projection) or if it is co-spatial with the disk but unable to alter the PAH grains.

\section{Conclusion} \label{sec:Conclusion}

Utilizing the exceptional resolution and sensitivity of \textit{JWST}, we investigate the nearby merging LIRG NGC 3256 on scales of $<$100 pc. Here, we showcase the capability of \textit{JWST} to spatially and spectrally resolve the MIR H$_2$ lines, allowing us to directly identify and trace the outflowing warm H$_2$ gas. Using the H$_2$ lines from 5.0--17.0 $\mu$m, we analyze the mass, temperature distribution, kinematics, and energetics of the outflowing warm H$_2$ gas, and assess the impact of the outflows on the local star formation. The following is a summary of our findings.


\textbullet\; Warm, outflowing H$_2$ gas is detected in a collimated outflow originating from the southern nucleus. The outflows extend out to a distance of 700 pc, and have a deprojected maximum velocity of $\sim$1,000 km s$^{-1}$. The emission is most intense near the far edges of the outflows, possibly due to impacting the disk of the northern galaxy. We do not detect any significant outflowing H$_2$ gas originating from the northern nucleus.

\textbullet\; Using the extensive wavelength coverage of \textit{JWST}, we use the full set of S(8)--S(1) H$_2$ lines to directly calculate the outflowing warm H$_2$ gas mass: $M\rm{_{warm,out}}$ = (1.4$\pm$0.2)$\times$10$^6\;M_{\odot}$, where the $M\rm{_{warm}}$/$M\rm{_{cold}}$ mass fraction is 6$\%$. With an outflow time scale of about $7\times10^5$ yr, the outflow mass rate is about $\dot{M}\rm{_{warm,out}}$ = 2.0$\pm$0.8 M$_{\odot}$ yr$^{-1}$. These masses yield an outflow energy of $E\rm{_{warm}}\;\sim\;8\times10^{54}$ erg and kinetic power of $P\rm{_{warm,out}}\;\sim\;4\times9^{41}$ erg s$^{-1}$. The mass loading factor of the warm outflowing gas is roughly unity. However, when taking into account the cold H$_2$ component, it could be five or higher. 

\textbullet\; The spatial map of the outflowing H$_2$ S(7)/S(1) flux ratio reveals higher ratios closer to the S nucleus. This indicates a larger fraction of warmer gas is closer to the nucleus and decreases as the outflow expands outwards, suggesting that the S nucleus is the heating source of the outflowing gas.

\textbullet\; Analysis of the gas excitation in the outflow regions using [\ion{Fe}{2}]/Pa$\beta$ and H$_2$/Br$\gamma$ line ratios shows enhanced ratios compared to those seen in typical star forming regions, indicating the outflow is energizing these regions, possibly as a shock. The value of the power-law index is also indicative of shocks, particularly in the blueshifted outflow. 

\textbullet\; Based on 6.2 $\mu$m PAH fluxes, we see no clear signs of the outflows significantly reducing the surface brightness of the PAH emission. Furthermore, analysis of the 6.2/7.7, and 11.3/7.7 PAH ratios do not show significant differences in the PAH ionization state and grain size between the outflow regions and those elsewhere in the disk.
\vspace{2mm}

In this article, we have showcased the unique capabilities of \textit{JWST} that enable us to explore the warm molecular gas in the MIR with both superb spectral and spatial resolutions. Further observations of other warm molecular outflows in the MIR and direct comparisons to the cold dense gas and the ionized atomic gas on similar physical scales will allow for a comprehensive study of the the multi-phase outflows in nearby star forming galaxies and AGN.\\

We thank the referee for their insightful feedback that helped improve the manuscript. The \textit{JWST} data presented in this paper were obtained from the Mikulski Archive for Space Telescopes (MAST) at the Space Telescope Science Institute, which is operated by the Association of Universities for Research in Astronomy, Inc., under NASA contract NAS 5-03127 for \textit{JWST}. These observations are associated with program 1328 and are supported by NASA grant ERS-01328. They can be accessed at \dataset[https://doi.org/10.17909/8mkr-xr82]{\doi{10.17909/8mkr-xr82}}. STScI is operated by the Association of Universities for Research in Astronomy, Inc., under NASA contract NAS5–26555. Support to MAST for these data is provided by the NASA Office of Space Science via grant NAG5–7584 and by other grants and contracts. T.B. and H.I. acknowledge support from JSPS KAKENHI Grant Number JP21H01129 and the Ito Foundation for Promotion of Science. A.S.E. and S.T.L. acknowledge support from NASA grants HST-GO15472 and HST-GO16914. V.U. acknowledges funding support from NSF Astronomy and Astrophysics Grant (AAG) No. AST-2408820, NASA Astrophysics Data Analysis Program (ADAP) grant No. 80NSSC23K0750, and STScI grant Nos. HST-AR-17063.005-A, HST-GO-17285.001-A, and JWST-GO-01717.001-A. C.R. acknowledges support from Fondecyt Regular grant 1230345 and ANID BASAL project FB210003 The Flatiron Institute is supported by the Simons Foundation. S.A. gratefully acknowledges support from ERC Advanced Grant 789410, the Swedish Research Council, and the Knut and Alice Wallenberg (KAW) foundation. F.M-S. acknowledges support from NASA through ADAP award 80NSSC19K1096. This paper makes use of the following ALMA data: ADS/JAO.ALMA$\#$2011.0.00525.S and 2015.1.00902.S. ALMA is a partnership of ESO (representing its member states), NSF (USA) and NINS (Japan), together with NRC (Canada), NSTC and ASIAA (Taiwan), and KASI (Republic of Korea), in cooperation with the Republic of Chile. The Joint ALMA Observatory is operated by ESO, AUI/NRAO and NAOJ.

\vspace{5mm}
\facilities{\textit{JWST}/NIRCam, \textit{JWST}/NIRSpec, \textit{JWST}/MIRI, \textsc{ALMA}, \textsc{ALMA Science Archive}, \textit{Gaia}, \textsc{Gaia ESA Archive}, \textsc{MAST}, \textsc{VLT}/SINFONI, \textsc{ESO Science Archive}}

\software{BADASS \citep{Sexton2021}, \textit{JWST} Science Calibration Pipeline \citep{Bushouse2023}, ASTROPY \citep{Astropy2013,Astropy2018,Astropy2022}}

\appendix

\begin{figure}[ht]
\gridline{\fig{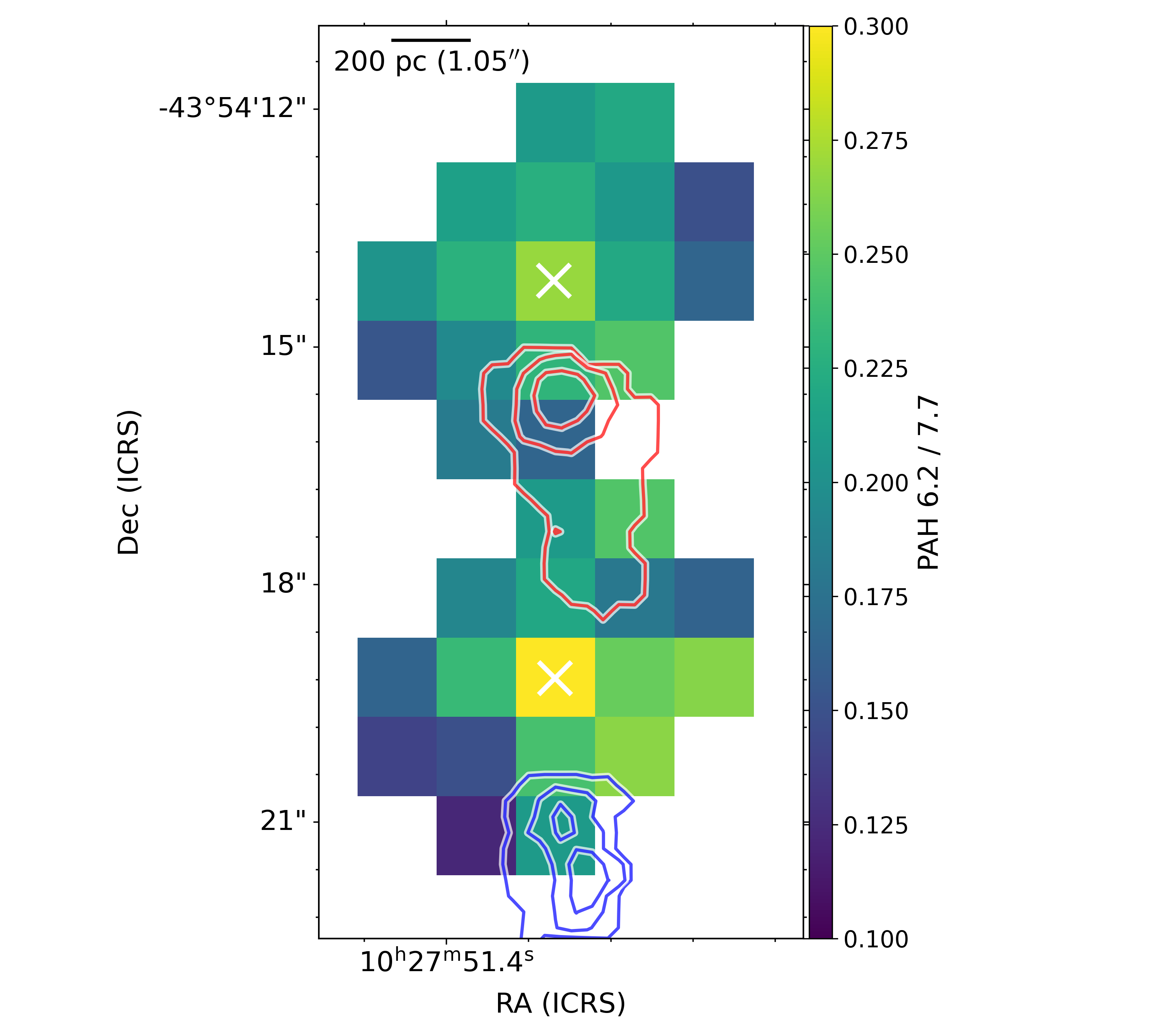}{0.5\textwidth}{}
\fig{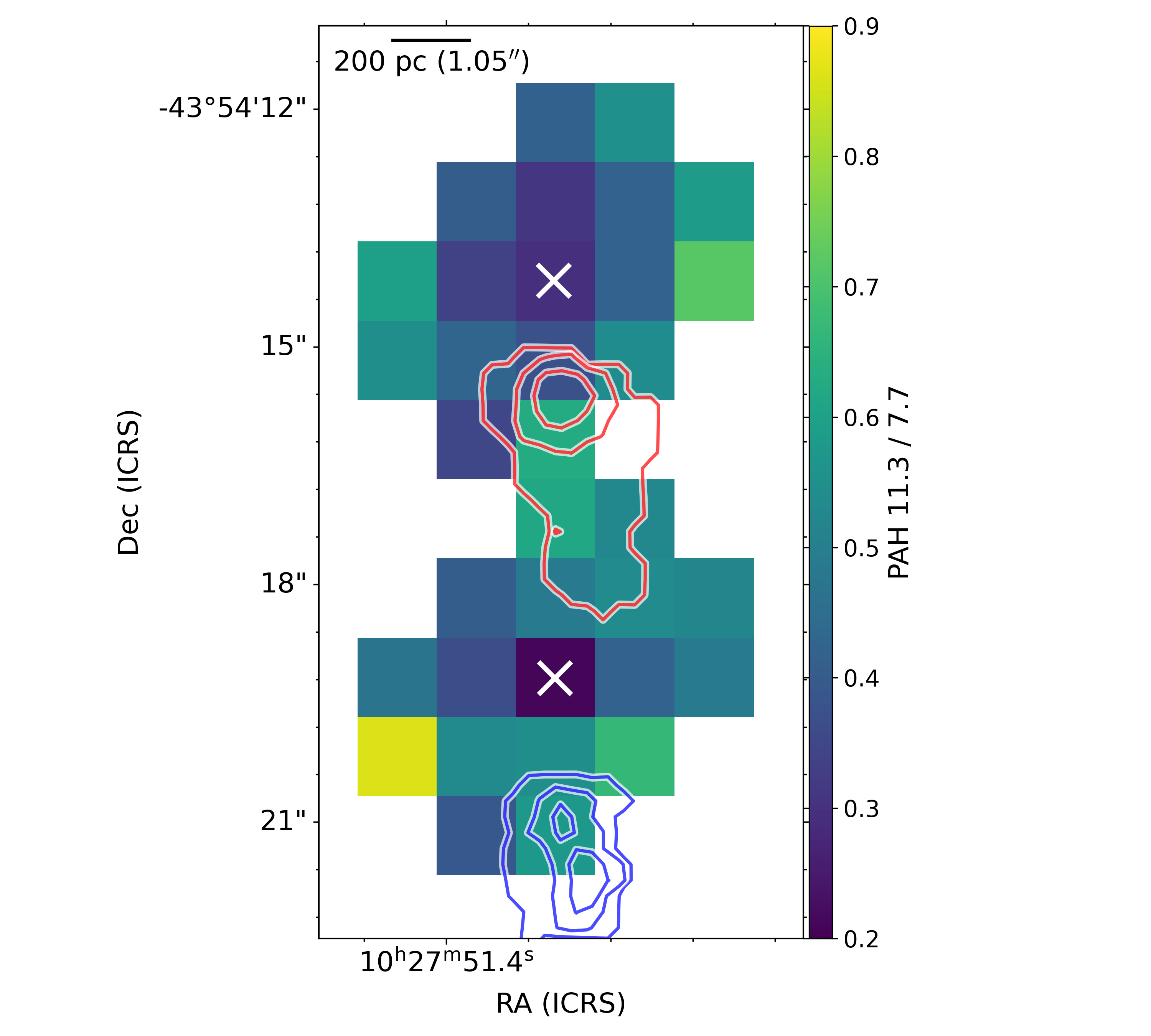}{0.5\textwidth}{}}
\caption{PAH$_{6.2}$/PAH$_{7.7}$ and PAH$_{11.3}$/PAH$_{7.7}$ ratio maps. Each grid is $1\farcs0\times1\farcs0$ in size. The locations of the nuclei are shown as white crosses and the red and blue contours show the locations of the outflows based on the H$_2$ S(1) 17.0 $\mu$m outflow emission.\label{fig:PAH_ratio_map}}
\end{figure}

The following section presents the \textsc{CRETA} grid extractions as well as the PAH$_{6.2}$/PAH$_{7.7}$ and PAH$_{11.3}$/PAH$_{7.7}$ ratio maps, as discussed in Section \ref{subsubsec:SF_feedback}. The grid sizes used to calculate the PAH ratios are $1\farcs0\times1\farcs0$, consistent with the PSF size at 11 $\mu$m, and are shown in Figure \ref{fig:PAH_ratio_map}. Two sets of grids were used, each centered on the nuclei, and they cover the full MIRI channel FoV. H$_2$ S(1) 17.0 $\mu$m outflow flux contours are overlaid to provide a sense of scale of each grid and help identify the ratios within the outflow regions. These ratios, along with those of the nuclei and disks, are listed in Table \ref{tab:PAH_ratios}. The mean PAH$_{6.2}$/PAH$_{7.7}$ and PAH$_{11.3}$/PAH$_{7.7}$ ratios of the outflow regions (0.21 and 0.54) are within the ranges of the disk regions (0.12 -- 0.27 and 0.30 -- 0.85), suggesting the outflows are not significantly altering the average PAH grain size nor their ionization state.

\begin{deluxetable}{ccc}
\setlength{\tabcolsep}{8pt}
\caption{PAH Flux Ratios} 
\label{tab:PAH_ratios}
\tablehead{\colhead{Region} & \colhead{PAH$_{6.2}$/PAH$_{7.7}$} & \colhead{PAH$_{11.3}$/PAH$_{7.7}$}}
\startdata
Northern Nucleus & 0.27$\pm$0.02 & 0.30$\pm$0.01\\
Southern Nucleus & 0.31$\pm$0.08 & 0.21$\pm$0.08\\
Red Outflow N & 0.17$\pm$0.03 & 0.63$\pm$0.03\\
 & 0.23$\pm$0.05 & 0.37$\pm$0.05\\
Red Outflow S & 0.22$\pm$0.04 & 0.49$\pm$0.02\\
 & 0.21$\pm$0.02 & 0.62$\pm$0.02\\
 & 0.18$\pm$0.05 & 0.54$\pm$0.05\\
 & 0.25$\pm$0.02 & 0.53$\pm$0.03\\
Blue Outflow & 0.21$\pm$0.02 & 0.57$\pm$0.02\\
Disk Regions & 0.12 -- 0.27 & 0.30 -- 0.85
\enddata
\tablecomments{Columns: (1) Locations of the extracted regions. (2-3) PAH$_{6.2}$/PAH$_{7.7}$ and PAH$_{11.3}$/PAH$_{7.7}$ flux ratios within each grid. For the disk regions, we list the range of ratio values.}
\end{deluxetable}

\pagebreak
\bibliography{Paper.bib}{}
\bibliographystyle{aasjournal}

\end{document}